\documentclass[aps,prb,amsmath]{revtex4}

\usepackage{graphicx}
\usepackage{bm}

\begin{document}

\title{Stress-driven instability in growing multilayer films}
\author{Zhi-Feng Huang}
 \email{huang@csit.fsu.edu}
 \altaffiliation[Present address: ]{School of Computational Science 
and Information Technology, Florida State University, Tallahassee, 
Florida 32306-4120}
\author{Rashmi C. Desai}
 \email{desai@physics.utoronto.ca}
\affiliation{%
Department of Physics, University of Toronto, 
Toronto, Ontario, Canada M5S 1A7
}%
\date{December 11, 2002 (revised); accepted for publication in Phys. Rev. B}

\begin{abstract}
We investigate the stress-driven morphological instability of 
epitaxially growing multilayer films, which are coherent and
dislocation-free. We construct a direct elastic analysis, from 
which we determine the elastic state of the system recursively 
in terms of that of the old states of the buried layers. In turn, 
we use the result for the elastic state to derive the morphological 
evolution equation of surface profile to first order of perturbations, 
with the solution explicitly expressed by the growth conditions and 
material parameters of all the deposited layers. We apply these 
results to two kinds of multilayer structures. One is the alternating 
tensile/compressive multilayer structure, for which we determine the 
effective stability properties, including the effect of varying 
surface mobility in different layers, its interplay with the global 
misfit of the multilayer film, and the influence of asymmetric 
structure of compressive and tensile layers on the system stability. 
The nature of the asymmetry properties found in stability diagrams 
is in agreement with experimental observations. The other multilayer 
structure that we study is one composed of stacked strained/spacer 
layers. We also calculate the kinetic critical thickness for the 
onset of morphological instability and obtain its reduction and 
saturation as number of deposited layers increases, which is 
consistent with recent experimental results. Compared to the 
single-layer film growth, the behavior of kinetic critical 
thickness shows deviations for upper strained layers.
\end{abstract}


\maketitle

\section{\label{sec:intro}Introduction}

In recent years, much attention has been focused on the 
heteroepitaxial growth of multilayer films for the purpose of 
fabricating high-quality low-dimensional structures. \cite{shchukin99}  
Currently, two kinds of multilayers are being actively studied: (i) 
the alternating tensile/compressive multilayers exploited as multiple 
quantum wells or short-period superlattices; \cite{norman98,ahrenkiel98,%
twesten99,norman02,ponchet93,ponchet94,ponchet95} and (ii) multisheets 
of coherently strained layers (or islands) alternately separated by 
spacer layers of substrate materials (strained/spacer arrays). 
\cite{shchukin99,tersoff96,teichert96,xie95,gonzalez00,nakata97,%
schmidt99,lethanh99,lethanh00,liu99,lafontaine98} Experimentally, these 
multilayer structures are grown coherently on the substrates, without 
the generation of undesirable misfit dislocations. When instability 
occurs and develops during the multilayer films growth, one of the 
most important phenomena is the formation of self-organized 
nanostructures. For the tensile/compressive multilayers, lateral 
modulations have been found during the epitaxial growth, showing as 
the modulated pattern of alternating vertical or columnar structures 
with strong contrast compositions. \cite{norman98,ahrenkiel98,%
twesten99,norman02,ponchet93,ponchet94} These modulation structures 
that form spontaneously are very useful for the quantum wires 
fabrication. For the strained/spacer multilayer structure, a 
remarkable phenomenon is the vertical self-organization of 
multisheet arrays of three-dimensional (3D) islands (quantum dots). 
\cite{shchukin99,tersoff96,teichert96,xie95,gonzalez00,nakata97,%
schmidt99,lethanh99,lethanh00,liu99} The islands are vertically correlated 
in successive strained layers, with the increasing uniformity 
(including island size and density) as the number of layers increases. 
These low-dimensional nanostructures (such as quantum wires and 
quantum dots), realized through the development of multilayer film 
instability, exhibit novel electronic and optical properties, 
including the band structure, carrier transport, etc., and are of 
intense interests both for fundamental studies and for application 
to devices with enhanced optoelectronic performance.

The mechanism of occurrence of instability in strained-layer 
growth is believed to be stress-driven. When a single-layer strained
film (with single component) is grown epitaxially on a substrate,
the morphological instability develops due to the nonuniform strain
energy distribution with respect to any surface corrugations and 
the surface diffusion processes driven by the difference of local
strain energy. \cite{asaro72,spencer91} When other layers composed 
of different materials, and consequently with different misfit 
strains with respect to the substrate, are deposited subsequently, 
as in the multilayer growth, the coupling of strain fields caused by 
buried layers and the growing one determines the morphological 
evolution of surface profile. This coupling is the source for 
self-organized growth of vertical islands array in the strained/spacer 
multilayer. \cite{tersoff96,xie95} For the tensile/compressive 
multilayer films, the phenomenon of lateral composition modulations 
can be explained by the morphological instability caused by the 
coupling strains, leading to the surface undulations and then to 
both the layer thickness modulations and composition modulations 
in the overall film through the stacking of layers. 
\cite{ponchet93,ponchet94}

Although in the coherent multilayer systems the processes of 
morphological evolution are mainly dominated by elastic effects, 
our understanding of the mechanisms is incomplete and not well 
developed due to the complexity of the multilayer structures. 
One of the main difficulties of the theoretical investigations is in 
determining the elastic fields of the system, especially the surface 
strain field caused by all the buried layers or islands. First, the 
system is inhomogeneous due to the stacking of layers with different 
materials; second, the geometry of this system, which is composed 
of top surface, finite thickness of film layers and the semi-infinite 
substrate, is not symmetric and is therefore different from the case 
of bulk materials. Moreover, the interfaces between different layers 
are generally not planar (except for the interfaces between strained 
and thick enough spacer layers), and then the system cannot be 
simplified to an equivalent and solvable semi-infinite solid as in 
the study of single-layer films. \cite{spencer91,guyer95,spencer00} 
Thus, it is very difficult to exactly calculate the elastic fields of 
the system.

The theoretical work for multilayer systems is rather sparse, and 
some approximations have to be used. For the tensile/compressive 
multilayer structure, Sridhar \textit{et al.} \cite{sridhar97} 
considered an infinite periodic and symmetric system with the 
dynamics controlled by interfacial diffusion, and focused on the 
interior layers that are far away from both surface and substrate, 
instead of the surface morphology that we are interested in this paper.  
In the recent work of Shilkrot \textit{et al.}, \cite{shilkrot00} 
the surface diffusion and morphological evolution is discussed, 
with the determination of stability for multilayer films. The Eshelby 
procedure \cite{eshelby57} is used for each buried interface, which 
is assumed to separately contribute to surface strain. Then the 
problem is approximated to be that of a misfit inclusion in a 
semi-infinite homogeneous isotropic elastic medium with the presence 
of a free surface. For a system of stacked coherent strained islands 
(strained/spacer multilayer), in the work of Tersoff \textit{et al.} 
\cite{tersoff96} each buried island is approximately treated as a 
point defect, and its contribution to surface strain is determined 
by using the continuum elasticity solutions of Maradudin and Wallis. 
\cite{maradudin80} Their results explain the vertical correlation 
between islands of successive layers, showing the self-organization 
of the system. The size and shape of stacked islands also exhibit 
the self-organized effects, as studied in a later work by Liu 
\textit{et al.}. \cite{liu99}

Here, we propose a recursive procedure and develop a continuum, 
dynamical model to directly describe the morphological evolution 
of these coherent multilayer structures during the epitaxial growth. 
Using the isotropic linear elasticity theory, we calculate in 
section \ref{sec:model} the elastic fields of the multilayer system 
\textit{directly} from the linearized boundary conditions at all 
the buried interfaces and the top surface. Based on our elastic 
analysis, the coupling and influence of all the buried layers and 
interfaces are determined in section \ref{sec:general} to describe 
the evolution of surface morphological perturbation. Within this 
framework, we investigate the stability properties of multilayer 
growing films by performing a linear analysis. Also, the effects of
different material parameters and growth conditions on morphological
stability/instability of multilayer films are determined and discussed.

We first apply in section \ref{sec:A/B} these results to the 
alternating tensile/compressive multilayer system, determining the 
effective stability properties. We investigate the effect of different 
surface mobilities of strained layers on the stability results, and 
the interplay between the relative mobility and the global misfit of 
multilayer film (especially the asymmetry between globally tensile 
and compressive films). In the absence of global strain, i.e., with 
a global strain-balanced condition, we find an interesting property 
that the system is more stable for the asymmetric structure (with 
respect to layer thickness and misfit) of compressive and tensile 
layers. The effects of relative surface tension and growth temperature 
are also studied.

Then in section \ref{sec:spacer} we investigate the strained/spacer 
multilayer structure (for the case of flat surface of spacer layers), 
and calculate the kinetic critical thickness for the onset of 3D 
instability. We obtain a reduction and saturation of critical 
thickness in the upper strained layers, well reproducing the 
recent experimental observations. We also find that the behavior 
of kinetic critical thickness for upper strained layers deviates 
from that of single-layer film growth for large misfit and small 
deposition rate.

\section{\label{sec:model}Model and elastic formulation}

Assume that a multilayer film is coherently deposited on a semi-infinite
substrate (denoted by index $0$) and consists of different layers $i=1$, 
$2$, $...$, $k$. In the general case, material of the deposited layer 
$i$ can be different from that of the substrate and that of all the 
other layers, and its misfit with respect to the substrate is denoted as 
$\epsilon_i$. As we know from the heteroepitaxial growth of a single-%
layer film, \cite{asaro72,spencer91,guyer95,spencer00,leonard98,huang02} 
the nonzero misfit strain can be relieved through the development of
morphological instability, leading to a nonplanar surface profile
$z=h_1(x,y)=l_1+\sum_{\bf q} \hat{h}_1({\bf q})\exp(iq_x x+iq_y y)$
for the first layer $1$, with a coherent interface at $z=h_0=\zeta(x,y)
=\sum_{\bf q} \hat{\zeta}({\bf q})\exp(iq_x x+iq_y y)$ (the average
position $\bar{\zeta}=0$) between film and substrate, where $l_1$ is 
the average thickness of first layer and $\hat{h}_1$ ($\hat{\zeta}$) 
denotes the surface (interface) morphological perturbation. Here we 
only consider the morphological modulations of the film, corresponding 
to the deposition of only a single component in each layer, and do not
consider the compositional evolution which is important in alloy growth 
systems and which results in more complicated properties due to its 
coupling with the morphological evolution. 
\cite{guyer95,spencer00,leonard98,huang02}

When the second and higher layers are deposited with different 
materials and thus with different misfits, the undulating surface 
is buried as a frozen rough interface if the growth temperature $T$ 
is not too high making interlayer diffusion negligible. Then the 
elastic state of the system is altered as a result of new deposition, 
and the morphological evolution of the new free surface is different 
from that of the first single layer due to the coupling of strain 
fields in different layers and the influence of buried interfaces on 
the elastic field and energy of the multilayer film as a whole. For 
a $k$-layer system, the growing surface profile is represented as 
$h_k(x,y,t)=\sum_{i=1}^{k-1} l_i + v_k t + \sum_{\bf q} \hat{h}_k
({\bf q},t)\exp(iq_x x+iq_y y)$, with $l_i$ ($1\leq i < k$) the 
average thickness of $i$th layer underneath, $v_k$ the deposition 
rate of top layer $k$, and $\hat{h}_k$ the top surface perturbation. 
When we consider the atomic diffusion along the top surface and the 
uniform deposition of material as the dominant mass-transport processes, 
the evolution equation of the surface morphological profile is given by
\begin{equation}
\frac{\partial h_k}{\partial t} = \Gamma_k \sqrt{g_k} \nabla_s^2
\frac{\delta {\cal F}}{\delta h_k} +v_k +\sqrt{g_k} \bm{\nabla \cdot
\eta}_c + \eta_k,
\label{eq-h}
\end{equation}
where $g_k=1+|\nabla h_k|^2$ denotes the determinant of surface metric
for top layer $k$, and $\Gamma_k=D_{s,k}N_{s,k}/k_B T N_{v,k}^2$ is 
the surface mobility with $D_{s,k}$ the surface diffusivity, $k_B$ the 
Boltzmann constant, $N_{s,k}$ and $N_{v,k}$ respectively the number 
density of atoms per unit surface area and per unit volume of layer $k$. 
Both the conserved thermal noise $\bm{\eta}_c$ and nonconserved deposition
noise $\eta_k$ are included in the above evolution equation, but are
neglected in the following studies since they do not alter the results
of linear analysis presented here. In Eq. (\ref{eq-h}), the free energy 
${\cal F}$ of the system consists of two terms:
\begin{equation}
{\cal F}=\gamma_k \int d^2r \sqrt{g_k} + \int_{-\infty}^{h_k} d^3r 
{\cal E}.
\label{eq-F}
\end{equation}
The first term is the surface energy with $\gamma_k$ the surface
tension at the top free surface of layer $k$, and tends to stabilize 
the film morphology. ${\cal E}$ in the second term represents the 
strain energy density, and is, in linear elasticity theory, 
${\cal E}=S_{\alpha\beta\xi\rho}\sigma_{\alpha\beta}\sigma_{\xi\rho}/2$ 
($\alpha$, $\beta$, $\xi$, $\rho=x$, $y$, $z$), where $\sigma_
{\alpha\beta}$ is the stress tensor. For an elastically isotropic 
system that we assume here, the elastic compliance tensor 
$S_{\alpha\beta\xi\rho}=\delta_{\alpha\xi}\delta_{\beta\rho}(1+\nu)/E-
\delta_{\alpha\beta}\delta_{\xi\rho}\nu/E$ with Young's modulus $E$ 
and Poisson ratio $\nu$. The form of total free energy in Eq. 
(\ref{eq-F}) implies that we have neglected the wetting effect of 
each deposited layer, which is not dominant when the layer thickness 
is assumed to be large enough. The wetting effect is related to the 
interaction between two successive layers or between film and substrate, 
and is important for thin films with thickness less than 3 ML. 
\cite{spencer97}

Substituting Eq. (\ref{eq-F}) into Eq. (\ref{eq-h}), we obtain the
expression for surface profile evolution of top growing layer $k$:
\begin{equation}
\partial h_k / \partial t = \Gamma_k \sqrt{g_k} \nabla_s^2
\left [ \gamma_k \kappa + {\cal E}_k(h_k,h_{k-1},...,\zeta) 
\right ] + v_k,
\label{eq-h_k}
\end{equation}
where $\kappa$ is the surface curvature, and ${\cal E}_k$ is the elastic 
energy density evaluated at surface $z=h_k$ and is dependent on all the
surface and interface morphologies due to the coupling of strain fields.
Since here we focus on the stability of the multilayer system and hence 
the early evolution regime with small perturbations, to determine the 
corresponding important properties we only need the \textit{first-order} 
evolution equation for surface morphology.
Expanding Eq. (\ref{eq-h_k}) in Fourier space to first order of the 
morphological perturbation, we obtain the linearized evolution equation 
that will be used in the following linear analysis:
\begin{equation}
\partial \hat{h}_k({\bf q},t) / \partial t = -\Gamma_k q^2
\left [ \gamma_k q^2 \hat{h}_k({\bf q},t) + 
\hat{\cal E}_k({\bf q}) \right ],
\label{eq-h_q}
\end{equation}
with $\hat{\cal E}_k ({\bf q})$ the first order elastic energy density 
evaluated at the growing surface position. For the buried interfaces
between layers $i+1$ and $i$ ($1 \leq i \leq k-1$), the morphology
$h_i(x,y)=\sum_{j=1}^{i} l_j + \sum_{\bf q} \hat{h}_i({\bf q})
\exp(iq_x x+iq_y y)$ as well as its perturbation $\hat{h}_i$ are 
also determined by Eqs. (\ref{eq-h_k}) and (\ref{eq-h_q}) during 
the growth of layer $i$, with the parameters of layer $k$ replaced 
by those of layer $i$; then the morphology at the top of the $i$th 
layer remains unchanged (with the evaluation at $t=l_i/v_i$) when 
the subsequent layers are deposited.

The above evolution form of the surface morphology is analogous to
that used for single-layer film growth,\cite{spencer91,guyer95,%
spencer00,leonard98,huang02} but the crucial difference is in the 
determination of elastic energy density ${\cal E}$, which is affected 
by all the underlying layers and the substrate, which are made up of 
different materials, in the inhomogeneous multilayer systems. To 
obtain the elastic free energy in linear elasticity theory, the 
stress tensor of each layer $i$ ($1 \leq i \leq k$ or substrate 
$i=0$ with $\epsilon_0=0$)
\begin{equation}
\sigma_{\alpha\beta,i} = 2\mu_i \left [\frac{\nu_i}{1-2\nu_i}
u_{\rho\rho,i}\delta_{\alpha\beta}+u_{\alpha\beta,i}
- \frac{1+\nu_i}{1-2\nu_i}\epsilon_i\delta_{\alpha\beta} \right ]
\label{eq-stress}
\end{equation}
($\alpha$, $\beta=x$, $y$, $z$) should be evaluated. Here $\mu_i=
E_i/2(1+\nu_i)$ is the shear modulus of layer $i$, with $E_i$ and 
$\nu_i$ the Young's modulus and Poisson ratio, and $u_{\alpha\beta,i}
=(\partial_{\alpha} u_{\beta,i} + \partial_{\beta} u_{\alpha,i})/2$
is the linear elastic strain tensor governed by the elastic displacement
field $u_{\alpha,i}$ in layer $i$.

The ideal way to determine the elastic state of the system is to
directly solve the mechanical equilibrium equation 
\begin{equation}
\partial_{\beta} \sigma_{\alpha\beta,i}=0,
\label{eq-equi}
\end{equation}
and then obtain the elastic fields $u_{\alpha,i}$ for all the coupling 
layers and substrate ($0 \leq i \leq k$) without any approximations.
Here we have assumed that in this nonequilibrium system of growing
multilayer film, the mechanical equilibrium (\ref{eq-equi}) occurs
instantaneously compared to the film morphological evolution process.
The corresponding boundary conditions can be obtained by the 
consideration of the negligible external pressure on top growing 
surface, leading to
\begin{equation}
\sigma_{\alpha\beta,k} n_{\beta,k} =0 \qquad {\rm at} 
\quad z=h_k(x,y,t), 
\label{eq-bound1}
\end{equation}
and the coherence at the buried $(i+1)$/$i$ interface for $0\leq i
\leq k-1$:
\begin{equation}
\sigma_{\alpha\beta,i+1} n_{\beta,i} =\sigma_{\alpha\beta,i} n_{\beta,i}
\qquad {\rm and} \qquad u_{\alpha,i+1} = u_{\alpha,i}
\qquad {\rm at} \quad z=h_i(x,y),
\label{eq-bound2}
\end{equation}
with $n_{\beta,k}$ and $n_{\beta,i}$ the unit vectors normal to the 
surface of layer $k$ and $(i+1)$/$i$ interface, respectively. Also, 
in the semi-infinite substrate, the decay of strain field far away 
from the film results in
\begin{equation}
u_{\alpha,0}\rightarrow 0 \qquad {\rm for} \quad z\rightarrow-\infty.
\label{eq-bound3}
\end{equation}
Therefore, for a $k$-layers system with one top surface and $k$ 
interfaces, we have to solve $6(k+1)$ equations from boundary 
conditions (\ref{eq-bound1})--(\ref{eq-bound3}) to determine the
solution of mechanical equilibrium equation (\ref{eq-equi}) for all
the layers and substrate. Also, due to the \textit{nonplanar} 
interfaces between different layers, even if the elastic constants 
of all the layers and substrate are assumed to be identical, the
stability results of the system here are not equivalent to those of
a simplified semi-infinite solid as in the former study of single
layer film with flat film-substrate interface. \cite{spencer91,%
guyer95,spencer00} So to obtain the exact and direct solution of 
these equations is very difficult for large $k$, even in a linear 
analysis.

Here we propose a recursive procedure to directly calculate
the strain fields of this multilayer system to first order of
perturbations. In the framework of isotropic elasticity and the
linear analysis, the elastic free energy can be determined from the 
exact solution of the (first-order) elastic fields and then used in
dynamical equation (\ref{eq-h_q}) for the analysis of morphological
evolution.

The basic state of this multilayer structure corresponds to a 
uniform film with a planar growing surface at constant rate $v_k$
and all the planar and coherent interfaces underneath. Since the 
basic state of the substrate is stress-free, with $\bar{u}_{\alpha,0}
=\bar{u}_{\alpha\beta,0}=\bar{\sigma}_{\alpha\beta,0}=0$ ($\alpha$, 
$\beta=x$, $y$, $z$), and all the surface and interfaces are flat, 
due to the coherency of the film there are no in-plane displacement 
and strain for all the layers, so that $\bar{u}_{x,i}=\bar{u}_{y,i}
=0$ and $\bar{u}_{x\beta,i}=\bar{u}_{y\beta,i}=0$ ($\beta=x$, $y$, 
$z$ and $1 \leq i \leq k$). Also, from the boundary condition 
(\ref{eq-bound1}) on the top surface and the coherency at all the 
flat interfaces, we have $\bar{\sigma}_{\alpha z,i}=0$ with $\alpha=x$,
$y$, $z$ in each layer $i$. The displacement in the $z$ direction is
nonzero: $\bar{u}_{z,i}=\bar{u}_i (z - \sum_{j=1}^{i-1} l_j) +
\sum_{j=1}^{i-1}\bar{u}_j l_j$ with $\bar{u}_i=\epsilon_i(1+\nu_i)
/(1-\nu_i)$, and the Poisson relaxation in the $z$ direction leads
to constant strain $\bar{u}_{zz,i}=\bar{u}_i$. The basic-state film
is stressed in the lateral direction, with stresses $\bar{\sigma}
_{xx,i}=\bar{\sigma}_{yy,i}=\bar{\sigma}_i=-2\mu_i\bar{u}_i$ for
all the layers. Note that in general case, the elastic constants
used here ($E_i$, $\nu_i$, and $\mu_i$) can be different in different
layers and substrate.

The expansion form of the mechanical equilibrium equation for each
layer or substrate can be obtained by substituting Eq. (\ref{eq-stress}) 
into Eq. (\ref{eq-equi}) and expanding in Fourier space with respect to
perturbations around the above basic state (in the following expressions 
all the hatted quantities denote the corresponding Fourier perturbations):
\begin{equation}
(1-2\nu_i)(\partial_z^2 -q^2) \left [
\begin{array}{c}
\hat{u}_{x,i} \\ \hat{u}_{y,i} \\ \hat{u}_{z,i}
\end{array} \right ] + \left [
\begin{array}{c}
iq_x \\ iq_y \\ \partial_z
\end{array} \right ] 
\left ( iq_x \hat{u}_{x,i} + iq_y \hat{u}_{y,i} + 
\partial_z \hat{u}_{z,i} \right )=0.
\label{eq-equi-q}
\end{equation}
for $i$th layer ($1 \leq i \leq k$) or substrate ($i=0$), with general 
solution
\begin{eqnarray}
\left [ \begin{array}{c}
\hat{u}_{x,i} \\ \hat{u}_{y,i} \\ \hat{u}_{z,i}
\end{array} \right ] &=& \left [
\begin{array}{c}
\alpha_{x,i} \\ \alpha_{y,i} \\ \alpha_{z,i} 
\end{array} \right ] \cosh(qz) + \left [
\begin{array}{c}
\beta_{x,i} \\ \beta_{y,i} \\ \beta_{z,i}
\end{array} \right ] \sinh(qz) \nonumber\\
&-& \left [
\begin{array}{c}
C_i iq_x/q \\ C_i iq_y/q \\ D_i
\end{array} \right ] z \sinh(qz) - \left [
\begin{array}{c}
D_i iq_x/q \\ D_i iq_y/q \\ C_i 
\end{array} \right ] z \cosh(qz)
\label{eq-u-q}
\end{eqnarray}
for $1 \leq i \leq k$, and
\begin{equation}
\left [ \begin{array}{c}
\hat{u}_{x,0} \\ \hat{u}_{y,0} \\ \hat{u}_{z,0}
\end{array} \right ] = \left [
\begin{array}{c}
\alpha_{x,0} \\ \alpha_{y,0} \\ \alpha_{z,0}
\end{array} \right ] e^{qz} - \left [
\begin{array}{c}
iq_x/q \\ iq_y/q \\ 1
\end{array} \right ] B ze^{qz}
\label{eq-u0-q}
\end{equation}
for substrate $i=0$ due to boundary condition (\ref{eq-bound3}),
where the parameters $C_i=(iq_x \alpha_{x,i} + iq_y \alpha_{y,i}
+q \beta_{z,i})/(3-4\nu_i)$, $D_i=(iq_x \beta_{x,i} + iq_y \beta_{y,i}
+q \alpha_{z,i})/(3-4\nu_i)$, and $B=(iq_x \alpha_{x,0} + iq_y 
\alpha_{y,0} + q \alpha_{z,0})/(3-4\nu_0)$. These forms of differential 
equations and general solutions are the same as those for single-layer 
film, \cite{spencer91} but with different \textit{linearized} boundary 
conditions. From Eq. (\ref{eq-bound1}), we have
\begin{equation}
\hat{\sigma}_{\beta z,k}=iq_{\beta} \bar{\sigma}_k \hat{h}_k 
\quad (\beta=x, y) \qquad {\rm and} \qquad \hat{\sigma}_{zz,k}= 0,
\label{eq-bound1-q}
\end{equation}
at top surface $z=\bar{h}_k=\sum_{i=1}^{k-1} l_i + v_k t$, while
at the interface between layers $i+1$ and $i$, i.e., at
$z=\bar{h}_i=\sum_{j=1}^{i} l_j$ ($0 \leq i \leq k-1$), boundary
condition (\ref{eq-bound2}) leads to
\begin{subequations}
\label{eq-bound2-q}
\begin{eqnarray}
&& -iq_{\beta} (\bar{\sigma}_{i+1} - \bar{\sigma}_i) \hat{h}_i
+\hat{\sigma}_{\beta z,i+1} = \hat{\sigma}_{\beta z,i}
\quad (\beta=x, y), \nonumber\\
&& \hat{\sigma}_{zz,i+1}=\hat{\sigma}_{zz,i},
\label{eq-bound2a-q}
\end{eqnarray}
and
\begin{eqnarray}
&& \hat{u}_{\beta,i+1} = \hat{u}_{\beta,i}
\quad (\beta=x, y), \nonumber\\
&& (\bar{u}_{i+1}-\bar{u}_i) \hat{h}_i +\hat{u}_{z,i+1}
=\hat{u}_{z,i}.
\label{eq-bound2b-q}
\end{eqnarray}
\end{subequations}
Note that when $i=0$, we have $\bar{u}_0=\bar{\sigma}_0=0$ and 
$\hat{h}_0=\hat{\zeta}$ for the substrate, and condition 
(\ref{eq-bound3}) has already been used to obtain the solution 
form (\ref{eq-u0-q}).

All the above mechanical equilibrium equations and boundary conditions
are for the $k$-layer system, and are also fulfilled by the 
$(k-1)$-layer system before layer $k$ is deposited, with $k$ replaced
by $k-1$. Let us denote the elastic displacement field for this
$(k-1)$-layer system (with average film thickness $\sum_{i=1}^{k-1}l_i$)
by $u_{\alpha,i}|_{k-1}$ with $0 \leq i \leq k-1$ and $\alpha=x$, $y$, 
$z$, where ``$|_{k-1}$'' represents that the system is of $(k-1)$ layers. 
When a new layer $k$ is deposited, the elastic state of the whole 
multilayer is changed and the displacement field is now $u_{\alpha,i}|_k$ 
(for the $k$-layer system with average film thickness $\sum_{i=1}^{k-1}
l_i + v_k t$). For all the buried $(k-1)$ layers and substrate, this
elastic field is composed of two contributions:
\begin{equation}
u_{\alpha,i}|_k = u^o_{\alpha,i}|_k + u^n_{\alpha,i}|_k \qquad
{\rm for} \quad 0 \leq i \leq k-1.
\label{eq-u}
\end{equation}
Here the first contribution $u^o_{\alpha,i}|_k$ is the \textit{old-state} 
field before the deposition of new layer $k$, and is identical to 
$u_{\alpha,i}|_{k-1}$ determined by the previous $(k-1)$-layer system
and assumed to be known. The second contribution $u^n_{\alpha,i}|_k$ 
represents all the change of strain field caused by the new deposition, 
and is the unknown \textit{new-state} field. The other unknown quantity 
is the elastic field of the new top layer $k$: $u_{\alpha,k}|_k$, and 
both of them should be calculated from the mechanical equilibrium 
equations and boundary conditions.

Since the old-state field $u^o_{\alpha,i}|_k=u_{\alpha,i}|_{k-1}$ 
obeys the mechanical equilibrium equation (\ref{eq-equi}) (or 
(\ref{eq-equi-q})), from Eq. (\ref{eq-u}) we know that new-state 
elastic field $u^n_{\alpha,i}|_k$ also follows the same equation, 
with the same solution forms (\ref{eq-u-q}) and (\ref{eq-u0-q}), and 
so does the unknown elastic field $u_{\alpha,k}|_k$ of the new layer 
$k$. For the linearized boundary conditions (\ref{eq-bound1-q}) and 
(\ref{eq-bound2-q}), when we replace $\hat{u}_{\alpha,i}$ by formula 
(\ref{eq-u}), $\hat{\sigma}_{\alpha\beta,i}$ by $\hat{\sigma}_
{\alpha\beta,i}|_k=\hat{\sigma}^o_{\alpha\beta,i}|_k + \hat{\sigma}^n
_{\alpha\beta,i}|_k$ ($0 \leq i \leq k-1$) which can be obtained 
according to Eq. (\ref{eq-stress}), and use the similar boundary 
conditions for $(k-1)$-layer system, which are: at $z=\bar{h}_{k-1}$,
\[ 
\hat{\sigma}_{\beta z,k-1}|_{k-1}=iq_{\beta} \bar{\sigma}_{k-1} 
\hat{h}_{k-1} \quad {\rm and} \quad \hat{\sigma}_{zz,k-1}|_{k-1}= 0, 
\]
and at $z=\bar{h}_i$ for $(i+1)$/$i$ interface ($0 \leq i \leq k-2$),
\begin{eqnarray*}
& -iq_{\beta} (\bar{\sigma}_{i+1} - \bar{\sigma}_i) \hat{h}_i
+\hat{\sigma}_{\beta z,i+1}|_{k-1} = \hat{\sigma}_{\beta z,i}|_{k-1}
\quad {\rm and} \quad 
\hat{\sigma}_{zz,i+1}|_{k-1}=\hat{\sigma}_{zz,i}|_{k-1},& \\
& \hat{u}_{\beta,i+1}|_{k-1} = \hat{u}_{\beta,i}|_{k-1}
\quad {\rm and} \quad 
(\bar{u}_{i+1}-\bar{u}_i) \hat{h}_i +\hat{u}_{z,i+1}|_{k-1}
=\hat{u}_{z,i}|_{k-1},&
\end{eqnarray*}
we can derive the simplified boundary conditions for new-state fields 
$u^n_{\alpha,i}|_k$ ($0 \leq i \leq k-1$) and  for the top-layer field
$u_{\alpha,k}|_k$. On top surface of the $k$-layer system, i.e., 
$z=\bar{h}_k$, condition (\ref{eq-bound1-q}) remains unchanged:
\begin{equation}
\hat{\sigma}_{\beta z,k}|_k=iq_{\beta} \bar{\sigma}_k \hat{h}_k 
\quad (\beta=x, y) \qquad {\rm and} \qquad \hat{\sigma}_{zz,k}|_k= 0,
\label{eq-bound1-n}
\end{equation}
while at $z=\bar{h}_{k-1}$ for the $k$/$(k-1)$ interface, we have (with 
$\beta=x$, $y$ and $u_{\alpha,k-1}|_{k-1}=u^o_{\alpha,k-1}|_k$ for 
$\alpha=x$, $y$, $z$)
\begin{subequations}
\label{eq-bound2-n}
\begin{eqnarray}
& -iq_{\beta} \bar{\sigma}_k \hat{h}_{k-1} + \hat{\sigma}
_{\beta z,k}|_k = \hat{\sigma}^n_{\beta z,k-1}|_k 
\quad {\rm and} \quad \hat{\sigma}_{zz,k}|_k 
= \hat{\sigma}^n_{zz,k-1}|_k,& \label{eq-bound2a-n}\\
& \hat{u}_{\beta,k}|_k = \hat{u}_{\beta,k-1}|_{k-1}
+ \hat{u}^n_{\beta,k-1}|_k  \quad {\rm and}& \nonumber\\
& (\bar{u}_k-\bar{u}_{k-1}) \hat{h}_{k-1} +\hat{u}_{z,k}|_k
=\hat{u}_{z,k-1}|_{k-1}+ \hat{u}^n_{z,k-1}|_k.&
\label{eq-bound2b-n}
\end{eqnarray}
\end{subequations}
At all the other underlying interfaces, the boundary conditions are 
much simpler:
\begin{equation}
\hat{\sigma}^n_{\alpha z,i+1}|_k = \hat{\sigma}^n_{\alpha z,i}|_k
\quad {\rm and} \quad 
\hat{u}^n_{\alpha,i+1}|_k=\hat{u}^n_{\alpha,i}|_k
\label{eq-bound3-n}
\end{equation}
at $z=\bar{h}_i$ with $\alpha=x$, $y$, $z$ and $0 \leq i \leq k-2$.

\begin{figure}
\resizebox{12.5cm}{!}{\includegraphics{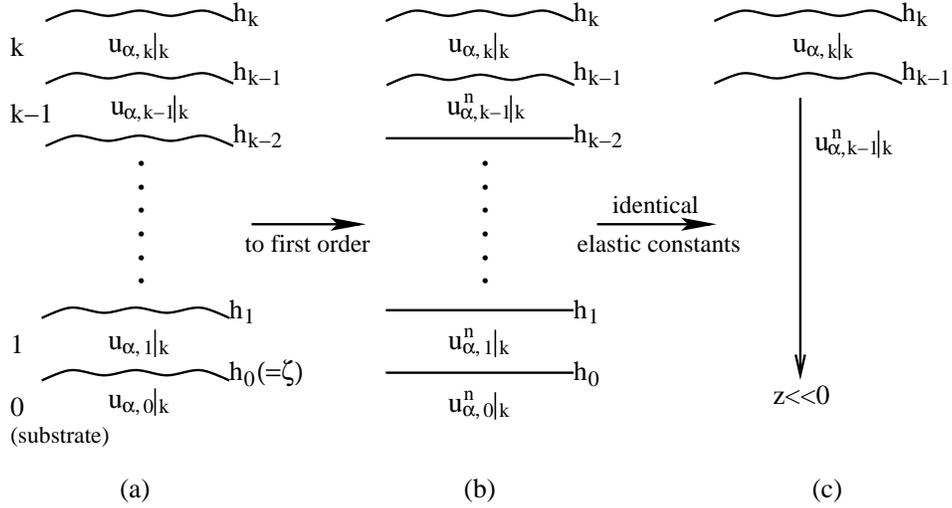}}
\caption{\label{fig-sche}Schematic diagram of the simplification
procedure for determining the elastic state of the multilayer system.
(a) The original $k$-layer structure that we study, with growing and
undulating surface (located at $z=h_k(x,y,t)$) and with nonplanar and 
frozen interfaces (located at $z=h_i(x,y)$) between buried layers 
$i=k-1$, $k-2$, $...$, $1$, and $0$ (substrate). The elastic 
displacement field of each layer is indicated as $u_{\alpha,i}|_k$ 
with $\alpha=x$, $y$, $z$. (b) The effective system of which the 
elastic state is identical to that of system (a) to first order 
of perturbations. Note the planar interfaces between layers $i$
and $i-1$ with $0 \leq i \leq k-1$, which are considered to be in the 
effective elastic media of new-state fields $u^n_{\alpha,i}|_k$.
(c) The simplified system that is equivalent to system (b) when
the elastic constants of all the layers and substrates are assumed
to be identical. Then we have only one layer on a semi-infinite
effective medium, with a rough surface and single nonplanar interface.
In both (b) and (c), top free surface obeys the boundary condition
(\ref{eq-bound1-n}) and the $k$/$(k-1)$ interface fulfills the 
special boundary condition (\ref{eq-bound2-n}).}
\end{figure}

Note that the boundary conditions (\ref{eq-bound3-n}) for all the
interfaces buried below layer $k-1$ are equivalent to those of 
\textit{planar} interfaces in the elastic media with displacement 
fields $\hat{u}^n_{\alpha,i}|_k$. Thus, from all the boundary 
conditions (\ref{eq-bound1-n})--(\ref{eq-bound3-n}) and the 
mechanical equilibrium equations followed by elastic fields $\hat{u}
_{\alpha,k}|_k$ and $\hat{u}^n_{\alpha,i}|_k$, we can obtain an
effective elastic system for which the elastic state is identical
(to first order) to that of the $k$-layer system we are interested 
in. This effective system is composed of a growing and undulating 
surface of top layer $k$ with condition (\ref{eq-bound1-n}), 
a nonplanar interface between layers $k$ and $k-1$ with special
boundary conditions (\ref{eq-bound2-n}), as well as the effective
planar interfaces (with respect to the new-state fields $\hat{u}^n_
{\alpha,i}|_k$) between all the other buried layers and between 
film and substrate. The schematic illustration of this procedure
is shown in Fig. \ref{fig-sche} (a) and (b), where part (a) presents
the original $k$-layer system with nonplanar surface and interfaces
and with elastic fields $u_{\alpha,i}|_k$ ($0 \leq i \leq k$) to be 
determined, and part (b) is the effective system (related to elastic 
fields $u_{\alpha,k}|_k$ for top layer $k$ and $u^n_{\alpha,i}|_k$ 
for all layers underneath) that we obtain here to first order of 
perturbations.

This elastic system can be further simplified and then solved exactly
when we consider the identical elastic constants for all the layers and 
substrate. The assumption of identical elastic constants is applicable
to the system of which substrate and all the layers in the film are
composed of materials with similar elastic constants, and is used in
the following calculations so that we could focus on the effects related
to surface mobilities and layer thicknesses. Note that for a single-layer
film, grown on a flat substrate, the elastic behavior of the system is 
equivalent to that of a semi-infinite stressed solid in the case of 
equal elastic constants. \cite{spencer91,spencer00} This can also be
obtained by combining the elastic solution forms in Eq. (\ref{eq-u-q}) 
for layer $1$ and in Eq. (\ref{eq-u0-q}) for substrate with the flat 
boundary condition (\ref{eq-bound3-n}), which leads to the identical 
(new-state) elastic fields for the first layer and substrate, i.e., 
$\hat{u}^n_{\alpha,1}|_k=\hat{u}^n_{\alpha,0}|_k$. Similarly, from
all the effectively planar boundaries below layer $k-1$ (as shown in 
Eq. (\ref{eq-bound3-n}) and Fig. \ref{fig-sche} (b)), we can obtain 
the same solution for the new-state elastic fields in all the buried
layers and substrate: $\hat{u}^n_{\alpha,0}|_k=\hat{u}^n_{\alpha,1}|_k
=...=\hat{u}^n_{\alpha,k-1}|_k$. This implies that the elastic state
(with respect to $ \hat{u}^n_{\alpha,i}|_k$) for all these layers and
substrate ($0 \leq i \leq k-1$) is equivalent to that of a single 
semi-infinite effective medium. 

Consequently, under the approximation of identical elastic constants 
and in a first order analysis, we can simplify the original $k$-layer 
structure to an elastically equivalent system composed of a single 
layer $k$ (with elastic field $u_{\alpha,k}|_k$) and an underlying 
effective semi-infinite elastic medium (with respect to field 
$u^n_{\alpha,k-1}|_k$), as illustrated in Fig. \ref{fig-sche} (c). 
Now we only have one nonplanar interface at $z=h_{k-1}$, but 
\textit{with special boundary condition (\ref{eq-bound2-n}) which 
is different from that of the usual single-layer film growth}. The 
undulating free surface at $z=h_k$ satisfies the condition 
(\ref{eq-bound1-n}) as usual. The elastic state of this effective 
system can be solved directly, in a manner similar to that for
the case of a single-layer film with rough film-substrate interface
(but with different boundary condition). \cite{huang02} Thus, from
Eq. (\ref{eq-u}) (and noting the boundary condition (\ref{eq-bound2-n})), 
we can obtain the elastic fields of the original $k$-layer system 
$\hat{u}_{\alpha,i}|_k$ ($0 \leq i \leq k$) expressed by that of the 
previously grown $(k-1)$-layer system $\hat{u}_{\alpha,j}|_{k-1}$ 
($0 \leq j \leq k-1$).

Using the same method, elastic fields $\hat{u}_{\alpha,j}|_{k-1}$ 
can be calculated in terms of $\hat{u}_{\alpha,j}|_{k-2}$ for the 
$(k-2)$-layer system. This procedure is repeated by expressing the 
elastic fields for $i$-layer system in terms of those for $(i-1)$-layer 
system with $i=k$, $k-1$, $k-2$, $...$, $2$, until we reach the 
single-layer system for which the elastic fields $\hat{u}_{\alpha,i}|_1$
($i=0$, $1$) have already been obtained for both the cases of planar 
\cite{spencer91} and rough \cite{huang02} film-substrate interface.
Thus, from this recursive procedure we can derive the first-order
elastic fields of the $k$-layer growing system in the expression of
known material parameters, and then calculate the elastic free energy
for the determination of morphological evolution equation.

\section{\label{sec:general}Morphological evolution for general case}

In the linearized equation (\ref{eq-h_q}) for surface morphological 
evolution, the first-order elastic energy density is $\hat{\cal E}_k
({\bf q})=\bar{\sigma}_k (\hat{u}_{xx}|_k + \hat{u}_{yy}|_k)$ with 
strain tensors $\hat{u}_{xx}|_k$ and $\hat{u}_{yy}|_k$ evaluated at
top surface $z=\bar{h}_k$. For the case of identical elastic constants
($E$, $\mu$, and $\nu$) in all the layers and the substrate, the method 
in Sec. \ref{sec:model} (or the effective system in Fig. \ref{fig-sche} 
(c)) yields
\begin{eqnarray}
\hat{\cal E}_k({\bf q}) &=& -E'\epsilon_k^2 q \hat{h}_k
+ E' \left [ \epsilon_k^2 - \frac{\epsilon_k\epsilon_{k-1}}{2(1-\nu)}
q l_k \right ] e^{-q l_k} q\hat{h}_{k-1} \nonumber\\
&+& \frac{E}{1-\nu}\epsilon_k \left [ (q l_k-1) e^{-q l_k} 
\left ( iq_x \hat{u}_{x,k-1}|_{k-1} + iq_y \hat{u}_{y,k-1}|_{k-1}
\right )_{z=\bar{h}_{k-1}} \right. \nonumber\\
&-& \left. q l_k e^{-q l_k} 
\left ( q\hat{u}_{z,k-1}|_{k-1} \right )_{z=\bar{h}_{k-1}} \right ],
\label{eq-E}
\end{eqnarray}
where $E'=2E(1+\nu)/(1-\nu)$ and $l_k=v_k t$ is defined for the top 
growing layer $k$. Also $\hat{u}_{\alpha,k-1}|_{k-1}(z=\bar{h}_{k-1})$ 
is obtained from the solution of elastic fields according to the 
recursive procedure presented in Sec. \ref{sec:model}, with the result:
\begin{eqnarray*}
 \left ( iq_x \hat{u}_{x,k-1}|_{k-1} 
+ iq_y \hat{u}_{y,k-1}|_{k-1} \right )_{z=\bar{h}_{k-1}} &=& 2(1-\nu)
\Bigglb [ q\bar{u}_{k-1} \hat{h}_{k-1} \nonumber\\
&-& q \sum\limits_{j=1}^{k-2}
\left ( \bar{u}_{j+1}-\bar{u}_j \right ) e^{-q\sum\limits_{i=j+1}^{k-1}
l_i} \hat{h}_j - q\bar{u}_1 e^{-q\sum\limits_{i=1}^{k-1} l_i}
\hat{\zeta} \Biggrb ], 
\end{eqnarray*}
\[
 \left ( iq_x \hat{u}_{x,k-1}|_{k-1} + iq_y \hat{u}_{y,k-1}|_{k-1}
-q\hat{u}_{z,k-1}|_{k-1} \right )_{z=\bar{h}_{k-1}} = q \bar{u}_{k-1}
\hat{h}_{k-1},
\]
with $\bar{u}_j=\epsilon_j (1+\nu)/(1-\nu)$.

Substituting Eq. (\ref{eq-E}) into Eq. (\ref{eq-h_q}), we obtain
the evolution equation of surface morphological perturbation:
\begin{eqnarray}
\partial \hat{h}_k(q,t) / \partial t &=& \sigma_k 
\hat{h}_k(q,t) - \Gamma_k E' \epsilon_k q^3 e^{-qv_k t} 
\nonumber\\
&\times& \left [ \sum\limits_{j=1}^{k-1} \left ( \epsilon_{j+1}
-\epsilon_j \right ) e^{-q\sum\limits_{i=j+1}^{k-1} l_i} 
\hat{h}_j(q) + \epsilon_1 e^{-q\sum\limits_{i=1}^{k-1} l_i}
\hat{\zeta}(q) \right ],
\label{eq-h_k-q}
\end{eqnarray}
where
\begin{equation}
\sigma_k = \Gamma_k \left ( E' \epsilon_k^2 q^3 - \gamma_k q^4 
\right )
\label{eq-sig}
\end{equation}
corresponds to the (single-layer) perturbation growth rate without
the coupling of buried layers and interfaces, and the interface 
perturbations $\hat{h}_j$ and $\hat{\zeta}$ are time independent 
due to the absence of interdiffusion. In Eq. (\ref{eq-h_k-q}), the 
influence of elastic properties and morphologies of all buried layers 
on the evolution of the top surface profile is shown explicitly. 
Note that due to the exponential decay term $\exp(-q\sum_{i=j+1}^{k-1} 
l_i)$, there is much smaller influence from lower buried layers, as 
usually expected.

Note that the above linearized dynamical equation for surface
morphology is different from that in Ref. \onlinecite{shilkrot00}
by Shilkrot \textit{et al.} for a periodic structure composed of
alternating layers of materials $A$ and $B$ (see Eq. (10) there), 
which was also derived to first order of perturbations with the 
assumption of equal elastic constants within the multilayer 
structure. This is attributed to the different methods in the 
elastic analysis of the system. Shilkrot \textit{et al.} used 
the Eshelby procedure \cite{eshelby57} by assuming each buried 
interface as the interface of a misfitting inclusion in a 
semi-infinite homogeneous and isotropic elastic medium with 
free surface, and considered the contribution from each interface 
to be separate. Here we apply a more direct procedure to determine 
the elastic state, and find that for this highly inhomogeneous 
system, an underlying interface can only be treated as one 
between an \textit{effective} semi-infinite homogeneous medium
(with respect to a new-state elastic field $u^n_{\alpha}$) and 
another elastic medium (of top layer) with finite thickness and 
free top surface, under the condition of identical elastic constants
and linear analysis (see Fig. \ref{fig-sche} (c)). The explicit 
form of our evolution equation is simpler than that of Ref. 
\onlinecite{shilkrot00} and so is our solution as well as the 
corresponding recurrence relation as shown in Sec. \ref{sec:A/B} 
for periodic $A$/$B$ system.

Using the initial condition $\hat{h}_k(q,t=0)=\hat{h}_{k-1}
(q)$, we can easily solve the evolution equation (\ref{eq-h_k-q}),
with the result for the top surface morphology:
\begin{eqnarray}
\hat{h}_k(q,t) &=& \frac{1}{qv_k + \sigma_k} \Bigglb \{
\left [ \left ( qv_k + \Gamma_k (E'\epsilon_k\epsilon_{k-1}q^3
-\gamma_k q^4) \right ) e^{\sigma_k t} + \Gamma_k E' \epsilon_k
(\epsilon_k - \epsilon_{k-1})q^3 e^{-qv_k t} \right ] 
\hat{h}_{k-1} \nonumber\\
&+& \left ( e^{-qv_k t} - e^{\sigma_k t} \right ) \Gamma_k E' 
q^3 \epsilon_k \Bigglb [ \sum\limits_{j=1}^{k-2} (\epsilon_{j+1}
- \epsilon_j) e^{-q \sum\limits_{i=j+1}^{k-1} l_i} \hat{h}_j
-\epsilon_1 e^{-q \sum\limits_{i=1}^{k-1} l_i} \hat{\zeta}
\Biggrb ] \Biggrb \},
\label{eq-solu}
\end{eqnarray}
which is one of the central results of this paper. In Eq. (\ref{eq-solu}),
the morphologies of buried interfaces ($\hat{h}_j$, $j=1$, $...$, $k-1$)
remain unchanged during the evolution process of top layer, as a result 
of negligible bulk diffusion of the film, and can be determined from
the similar calculation for lower $j$-layer system. In the case of
nonzero misfits for all the deposited layers, (i.e., $\epsilon_i \neq 0$ 
for all $i=1,2,...,k$), Eq. (\ref{eq-solu}) can be reduced recursively 
through the expression of the surface or interface perturbation in terms 
of the perturbations of two lower successive interfaces, leading to the 
form
\begin{equation}
\left ( \begin{array}{c}
\hat{h}_k(q,t) \\ \hat{h}_{k-1}(q) \end{array} \right )
= {\bf L}_k \prod\limits_{j=3}^{k-1} \bar{\bf{L}}_j
\left ( \begin{array}{c}
\hat{h}_2(q) \\ \hat{h}_1(q) \end{array} \right ).
\label{eq-h_k-L}
\end{equation}
In Eq. (\ref{eq-h_k-L}), ${\bf L}_k$ and $\bar{\bf{L}}_j$ ($0 \leq j
\leq k-1$) are $2 \times 2$ matrices:
\begin{equation}
{\bf L}_k = \left ( \begin{array}{cc}
L_{k,1}(q,t) & L_{k,2}(q,t) \\
1 & 0 \end{array} \right ), \qquad
\bar{\bf{L}}_j = \left ( \begin{array}{cc}
\bar{L}_{j,1}(q) & \bar{L}_{j,2}(q) \\
1 & 0 \end{array} \right ),
\label{eq-L_kj}
\end{equation}
where
\begin{eqnarray}
L_{k,1}(q,t) &=& \frac{1}{qv_k + \sigma_k} \left \{
\left [ qv_k + \Gamma_k \left ( E'\epsilon_k \epsilon_{k-1} q^3 
-\gamma_k q^4 \right ) \right ] e^{\sigma_k t}
+ \Gamma_k E' \epsilon_k (\epsilon_k - \epsilon_{k-1}) q^3
e^{-q v_k t} \right \} \nonumber \\
&-& L_{k,2}(q,t)~ e^{-\sigma_{k-1}l_{k-1}/v_{k-1}}, \nonumber \\
L_{k,2}(q,t) &=& -\frac{qv_{k-1} + \sigma_{k-1}}{qv_k + \sigma_k}~
\frac{\Gamma_k\epsilon_k}{\Gamma_{k-1}\epsilon_{k-1}}~
\frac{e^{\sigma_{k-1}l_{k-1}/v_{k-1}-ql_{k-1}}}
{e^{-ql_{k-1}} - e^{\sigma_{k-1}l_{k-1}/v_{k-1}}}~
\left ( e^{-q v_k t} - e^{\sigma_k t} \right ),
\label{eq-L_k}
\end{eqnarray}
and $\bar{L}_{j,1}(q)$ and $\bar{L}_{j,2}(q)$ are obtained by replacing
$k$ with $j$ as well as $t$ with $l_j/v_j$ in $L_{k,1}(q,t)$ and 
$L_{k,2}(q,t)$, respectively. Results for first and second layer 
perturbations $\hat{h}_1$ and $\hat{h}_2$ are presented in the Appendix, 
for both rough ($\zeta\neq 0$) and planar ($\zeta=0$) film-substrate 
interfaces. Therefore, from Eqs. (\ref{eq-h_k-L})--(\ref{eq-L_k}) and
the results in the Appendix, we obtain the explicit expression for 
first-order surface morphological perturbation $\hat{h}_k(q,t)$ in 
terms of known material parameters, which can be directly used for 
stability analysis as shown in the following sections.

Note that the above solution results apply to the general case for
which growth parameters (including the layer thicknesses, misfits, 
surface mobilities, surface tensions, and deposition rate) of different 
layers can be different, and then apply to both the non-periodic and 
periodic multilayer structures. Also we have assumed the identical 
elastic constants in the whole multilayer film, and the nonzero misfit 
strains for all the layers. In the case of zero misfit $\epsilon_i=0$ 
for some spacer layers $i$ (corresponding to the strained/spacer 
multilayer structures), the solution forms are simpler and will be 
discussed in Sec. \ref{sec:spacer}.

\section{\label{sec:A/B}Tensile/compressive multilayers: periodic
$\bm{A/B}$ system}

Here we apply the results for morphological evolution, derived in the 
last section, to the periodic tensile/compressive multilayer structures 
that have been investigated in many experiments, such as InAs/AlAs on
InP(001), \cite{norman98,ahrenkiel98,twesten99,norman02} 
GaInP(InGaAs)/InAsP on InP(001), \cite{ponchet93,ponchet94,ponchet95}
etc. This type of multilayer film consists of two kinds of alternating 
layers with materials $A$ and $B$. Then we have misfit strain 
$\epsilon_A$, layer thickness $l_A$, surface mobility $\Gamma_A$,
surface tension $\gamma_A$, and deposition rate $v_A$ for odd-number
strained layers, and the corresponding parameters $\epsilon_B$, $l_B$, 
$\Gamma_B$, $\gamma_B$, and $v_B$ for even-number layers, with nonzero 
misfits for all the layers: $\epsilon_A \neq 0$ and $\epsilon_B \neq 0$.

\subsection{\label{sec:A/B-formul}Formulation of morphological evolution}

For this periodic $A$/$B$ system, the solution for the top surface
morphological perturbation can be directly obtained from Eqs. 
(\ref{eq-h_k-L})--(\ref{eq-L_k}). For a $k$-layer structure, when
$k$ is odd implying that the top layer $k$ is of material $A$, we have
\begin{equation}
\left ( \begin{array}{c}
\hat{h}^*_k(q^*,\tau) \\ \hat{h}^*_{k-1}(q^*) \end{array} \right )
= {\bf L}_A \left ( \bar{\bf{L}}_B \bar{\bf{L}}_A \right )
^{\frac{k-3}{2}}
\left ( \begin{array}{c}
\hat{h}^*_2(q^*) \\ \hat{h}^*_1(q^*) \end{array} \right )
\qquad (k \geq 3),
\label{eq-h_A}
\end{equation}
with $2 \times 2$ matrices:
\begin{equation}
{\bf L}_{A(B)} = \left ( \begin{array}{cc}
L_{A(B)1}(q^*,\tau) & L_{A(B)2}(q^*,\tau) \\
1 & 0 \end{array} \right ), \qquad
\bar{\bf{L}}_{A(B)} = \left ( \begin{array}{cc}
\bar{L}_{A(B)1}(q^*) & \bar{L}_{A(B)2}(q^*) \\
1 & 0 \end{array} \right ),
\label{eq-L_AB}
\end{equation}
while for even number of deposited layers, the top layer $k$ is of 
material $B$ and the solution form is 
\begin{equation}
\left ( \begin{array}{c}
\hat{h}^*_k(q^*,\tau) \\ \hat{h}^*_{k-1}(q^*) \end{array} \right )
= {\bf L}_B \bar{\bf{L}}_A \left ( \bar{\bf{L}}_B \bar{\bf{L}}_A 
\right )^{\frac{k-4}{2}}
\left ( \begin{array}{c}
\hat{h}^*_2(q^*) \\ \hat{h}^*_1(q^*) \end{array} \right )
\qquad (k \geq 4).
\label{eq-h_B}
\end{equation}
Here, the solutions (\ref{eq-h_A}) and (\ref{eq-h_B}) are expressed
in dimensionless form, and we have used the rescaling
\begin{equation}
q^*=q l_A^0, \qquad \tau=t/\tau_A^0, \qquad {\rm and} \qquad
\hat{h}^*_i = \hat{h}_i/l_A^0 \quad (0 \leq i \leq k), 
\label{eq-qth^*}
\end{equation}
with the characteristic length and time scale respectively given as: 
\begin{equation}
l_A^0 = \gamma_A/(E'\epsilon_A^2),
\label{eq-l_A^0}
\end{equation}
and 
\begin{equation}
\tau_A^0=\gamma_A^3/ (\Gamma_A E'^4 \epsilon_A^8).
\label{eq-tau_A^0}
\end{equation}
These are determined according to the (single-layer film) growth 
of layer $1$ (assumed of material $A$). We also introduce the 
characteristic velocity
\begin{equation}
v_A^0=l_A^0/\tau_A^0=\Gamma_A E'^3 \epsilon_A^6 / \gamma_A^2,
\label{eq-v_A^0}
\end{equation}
as well as the following transformations
\begin{eqnarray}
& l_A^*=l_A/l_A^0, \qquad l_B^*=l_B/l_A^0, & \nonumber\\
& v_A^*=v_A/v_A^0, \qquad v_B^*=v_B/v_A^0, & \nonumber\\
& \Gamma_B^*=\Gamma_B/\Gamma_A, \quad 
\epsilon_B^*=\epsilon_B/\epsilon_A, \quad
\gamma_B^*=\gamma_B/\gamma_A. &
\label{eq-trans}
\end{eqnarray}
Then the matrix elements in Eq. (\ref{eq-L_AB}) are rescaled as
\begin{eqnarray}
L_{A1}(q^*,\tau) &=& \frac{1}{q^*v_A^* + \sigma_A^*} \left \{
\left [ q^* v_A^* + \left ( \epsilon_B^* {q^*}^3 
-{q^*}^4 \right ) \right ] e^{\sigma_A^* \tau}
+ (1 - \epsilon_B^*) {q^*}^3 e^{-q^* v_A^* \tau} \right \}
\nonumber \\
&-& L_{A2}(q^*,\tau)~ e^{-(\sigma_B^* /v_B^*) l_B^*}, \nonumber \\
L_{A2}(q^*,\tau) &=& -\frac{q^* + \sigma_B^*/v_B^*}
{q^* v_A^* + \sigma_A^*}~\frac{1}{(\Gamma_B^*/v_B^*)\epsilon_B^*}~
\frac{e^{(\sigma_B^* /v_B^*) l_B^* - q^* l_B^*}}
{e^{-q^* l_B^*} - e^{(\sigma_B^* /v_B^*) l_B^*}}~
\left ( e^{-q^* v_A^* \tau} - e^{\sigma_A^* \tau} \right ),
 \nonumber \\
L_{B1}(q^*,\tau) &=& \frac{1}{q^* + \sigma_B^*/v_B^*} \left \{
\left [ q^* + \frac{\Gamma_B^*}{v_B^*} \left ( \epsilon_B^* {q^*}^3 
-\gamma_B^* {q^*}^4 \right ) \right ] e^{\sigma_B^* \tau}
+ \frac{\Gamma_B^*}{v_B^*} \epsilon_B^* (\epsilon_B^* -1) {q^*}^3 
e^{-q^* v_B^* \tau} \right \}
\nonumber \\
&-& L_{B2}(q^*,\tau)~ e^{-\sigma_A^* l_A^* /v_A^*}, \nonumber \\
L_{B2}(q^*,\tau) &=& -\frac{q^* v_A^* + \sigma_A^*}
{q^* + \sigma_B^*/v_B^*}~ \frac{\Gamma_B^*}{v_B^*} \epsilon_B^* 
~\frac{e^{\sigma_A^* l_A^* /v_A^* - q^* l_A^*}}
{e^{-q^* l_A^*} - e^{\sigma_A^* l_A^* /v_A^*}}~
\left ( e^{-q^* v_B^* \tau} - e^{\sigma_B^* \tau} \right ),
\label{eq-L^*}
\end{eqnarray}
with $\bar{L}_{A1}(q^*)$, $\bar{L}_{A2}(q^*)$, $\bar{L}_{B1}(q^*)$,
and $\bar{L}_{B2}(q^*)$ obtained by replacing $\tau$ with $l_A^*/v_A^*$
or $l_B^*/v_B^*$ in Eq. (\ref{eq-L^*}), and
\begin{eqnarray}
\sigma_A^* &=& {q^*}^3 - {q^*}^4, \nonumber\\
\sigma_B^* &=& \Gamma_B^* \left ( {\epsilon_B^*}^2 {q^*}^3 
- \gamma_B^* {q^*}^4 \right ).
\label{eq-sigma}
\end{eqnarray}

>From Eq. (\ref{eq-sigma}), we can have $\sigma_A^* >0$ or
$\sigma_B^*>0$ for a certain wave number $q^*$, and then the 
expressions in Eq. (\ref{eq-L^*}) show that $|L_{A1(2)}| \gg 1$
or $|L_{B1(2)}| \gg 1$ when time $\tau$ is very large. Consequently,
we can obtain from Eqs. (\ref{eq-h_A}) and (\ref{eq-h_B}) that
for a certain top surface layer, when the growth time is large
enough the amplitude of surface perturbation $\hat{h}$ will grow
rapidly with time. Thus, in principle this multilayer system is 
\textit{unstable}.

However, in real experiments the thickness of each layer ($l_A$
or $l_B$) is fixed, including the top layer, then the growth of 
this multilayer system could be effectively \textit{stable} or 
\textit{unstable}. \cite{shilkrot00,norman98,ahrenkiel98,twesten99,%
norman02} That is, when the film surface is flat after the growth 
of large number of strained layers ($k \gg 1$), i.e., the initial 
perturbation decays with further deposition, the system is 
effectively stable; otherwise, if the initial perturbation is 
amplified and grows in amplitude, the system is unstable. For the 
latter case, the film surface is observed to be very rough, with 
regular or irregular modulations. 

The stability properties can be calculated from Eqs. (\ref{eq-h_A}) 
and (\ref{eq-h_B}), where $\hat{h}_k^*(q^*,\tau)$, ${\bf L}_A$, 
and ${\bf L}_B$ should be replaced by $\hat{h}_k^*(q^*)$, 
$\bar{\bf L}_A$, and $\bar{\bf L}_B$, respectively, and then 
the top surface perturbation is mainly determined by the power 
of matrix $\bar{\bf L}=\bar{\bf L}_B\bar{\bf L}_A$, which has 
two eigenvalues $\lambda_1$ and $\lambda_2$ given by
\begin{equation}
\lambda_{1,2}(q^*) = \frac{1}{2} \left \{ \left ( \bar{L}_{A2}
+\bar{L}_{B2}+\bar{L}_{A1}\bar{L}_{B1} \right ) \pm
\left [ \left ( \bar{L}_{A2}+\bar{L}_{B2}+\bar{L}_{A1}\bar{L}_{B1} 
\right )^2 - 4\bar{L}_{A2}\bar{L}_{B2} \right ]^{1/2} \right \}.
\label{eq-lambda}
\end{equation}
Assuming that the corresponding two eigenvectors of matrix $\bar{\bf L}$ 
are linearly independent, we obtain the explicit expression for the
top surface morphological perturbation $\hat{h}_k^*(q^*)$:
\begin{eqnarray}
\hat{h}_k^* &=& \bar{L}_{A1} \frac{\lambda_1^{\frac{k-1}{2}}
-\lambda_2^{\frac{k-1}{2}}}{\lambda_1 - \lambda_2} \hat{h}_2^*
+ \bar{L}_{A2} \left ( -\bar{L}_{B2} \frac{\lambda_1^{\frac{k-3}{2}}
-\lambda_2^{\frac{k-3}{2}}}{\lambda_1 - \lambda_2} +
\frac{\lambda_1^{\frac{k-1}{2}}-\lambda_2^{\frac{k-1}{2}}}
{\lambda_1 - \lambda_2} \right ) \hat{h}_1^*
\quad (k~{\rm odd}), \nonumber\\
\hat{h}_k^* &=& \left ( \frac{\lambda_1^{\frac{k}{2}}
-\lambda_2^{\frac{k}{2}}}{\lambda_1 - \lambda_2} -\bar{L}_{A2}
\frac{\lambda_1^{\frac{k-2}{2}}-\lambda_2^{\frac{k-2}{2}}}
{\lambda_1 - \lambda_2} \right ) \hat{h}_2^* +
\bar{L}_{A2}\bar{L}_{B1} \frac{\lambda_1^{\frac{k-2}{2}}
-\lambda_2^{\frac{k-2}{2}}}{\lambda_1 - \lambda_2} \hat{h}_1^*
\quad (k~{\rm even}), 
\label{eq-h-lamb}
\end{eqnarray}
which can be easily proved to be proportional to $\lambda^{k/2}$
[$\lambda=\max(\lambda_1,\lambda_2)$] for real eigenvalues, or to
$|\lambda|^{k/2}$ ($|\lambda|=|\lambda_1|=|\lambda_2|$) for a complex 
eigenvalue pair, when the number of total layers $k \rightarrow \infty$. 
Thus, the effective stability of this multilayer system (defined for 
$k\gg 1$) is determined by the magnitude of eigenvalue $|\lambda|$. 
When the maximum $|\lambda|$ (with respect to wave number $q^*$) is 
larger than $1$, the surface perturbation grows, leading to the 
instability of the multilayer film; otherwise, if $|\lambda| \leq 1$ 
for all the $q^*$ modes, the system is stable due to the decay of 
surface perturbations. According to Eqs. (\ref{eq-L^*}) and 
(\ref{eq-lambda}), the value of $|\lambda|$, and then the stability 
properties of the system, depend on $6$ dimensionless parameters 
$\Gamma_B^*/v_B^*$, $\epsilon_B^*$, $v_A^*$, $l_A^*$, $l_B^*$, and 
$\gamma_B^*$ that are defined in Eq. (\ref{eq-trans}).

The expression for $\lambda$ remains unchanged with respect to the 
interchange of $A$ and $B$ layers, since the eigenvalues of matrices 
$\bar{\bf L}=\bar{\bf L}_B\bar{\bf L}_A$ and $\bar{\bf L}'=
\bar{\bf L}_A\bar{\bf L}_B$, with $\bar{\bf L}_{A(B)}$ defined in 
Eq. (\ref{eq-L_AB}), are the same. Thus, the stability properties
of multilayer structure are determined by $|\lambda|$ and
are identical in the $k \rightarrow \infty$ limit for the $A/B$ 
and $B/A$ stacking forms.

Note that the stability analysis presented here is for very large
$k \gg 1$, while for the morphology of each layer $i$ ($i=1$, $...$, 
$k$), or for the phase \cite{shilkrot00} between two successive 
strained layers, which depends on the sign of $\hat{h}_i^*/\hat{h}_
{i-1}^*$, the evolution Eqs. (\ref{eq-h_A}) and (\ref{eq-h_B}) [with 
the explicit expressions similar to Eq. (\ref{eq-h-lamb})] should be 
used. Although the stability properties for $k\gg 1$ are independent 
of the type of the first layer or the interchange of $A$ and $B$ 
layers in the structure, the detailed morphology 
of each layer $i$ may be influenced by whether the first deposited 
layer is tensile or compressive, as seen from the expressions 
(\ref{eq-h_A}), (\ref{eq-h_B}), and (\ref{eq-h-lamb}).

\subsection{\label{sec:A/B-results}Stability results}

In this subsection, we present the results of stability diagrams for 
various (rescaled) material parameters and growth conditions. The 
stability properties are determined through the maximum magnitude 
of eigenvalue $\lambda$, with the use of the above analytic 
expressions and discussions. One of the factors that we focus on 
is the surface mobilities of strained layers. 
During the growth process, different mobilities of $A$ and $B$ 
layers results in different interface morphologies which are frozen 
in due to the deposition of upper layers, but which can influence 
the system's strain field and the evolution of top surface profile.

A common feature of the stability diagrams presented in this 
subsection is that the multilayer system can be effectively 
stabilized by large enough deposition rate ($v_A^*$ and $v_B^*$), 
as also found in experiments. \cite{ponchet95,ahrenkiel98}
This is similar to single-layer film growth: \cite{spencer91,%
guyer95,spencer00,leonard98,huang02} for large enough deposition rate, 
the morphological perturbation of layer surface does not have enough 
time to develop through surface diffusion, since it is buried and
frozen by fast subsequent materials deposition.

\subsubsection{\label{sec:Gamma}Effect of different surface mobilities
in strained layers}

\begin{figure}
\resizebox{8.cm}{!}{\includegraphics{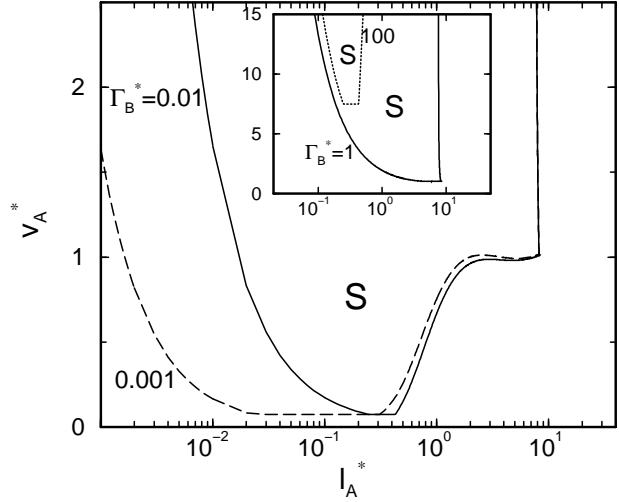}}
\caption{\label{fig-v-l}Stability diagram of rescaled layer thickness
$l_A^*$ versus deposition rate $v_A^*$, with parameters $\epsilon_B^*
=-1$, $l_B^*/l_A^*=1$, $v_B^*/v_A^*=1$, and $\gamma_B^*=1$ for the 
growing multilayer system. Different relative surface mobilities 
$\Gamma_B^*=0.001$ (dashed curve) and $0.01$ (solid curve) are used, 
as indicated in the figure. Region marked as ``$S$'' is effectively 
stable. Inset: $\Gamma_B^*=1$ (solid) and $100$ (dotted).}
\end{figure}

Stability diagrams in the $l_A^*$-$v_A^*$ space are shown in 
Fig. \ref{fig-v-l} for different relative surface mobilities
$\Gamma_B^*=\Gamma_B/\Gamma_A=0.001$, $0.01$, $1$, and $100$ 
(other parameters are $\epsilon_B^*=-1$, $l_A^*=l_B^*$, 
$v_A^*=v_B^*$, and $\gamma_B^*=1$). Important features are:
larger deposition rate imply greater stability, very thick or very
thin layers imply instability and for large $l_A$ and $l_B$, multilayer
system loses coherency; larger relative mobility $\Gamma_B^*$ leads to
more instability (note the larger unit-scale for $v_A^*$ in the inset).

To understand the effect due to 
different mobilities, we consider that the increase of $\Gamma_B^*=
\Gamma_B/\Gamma_A$ corresponds to keeping the parameters of material 
$A$ (including $\Gamma_A$, $\epsilon_A$, and $\gamma_A$) unchanged, 
as seen in the rescaling Eq. (\ref{eq-trans}), and to the increase of 
surface mobility $\Gamma_B$ of layer $B$. Consequently, for larger 
$\Gamma_B^*$, the morphological instability is enhanced due to a 
faster diffusion on surface of layer $B$ (driven by nonuniform strain 
energy distribution \cite{asaro72,spencer91}), and then larger 
deposition rates $v_A^*$ and $v_B^*$ are needed to suppress it. In 
multilayer experiments, different mobilities result in different 
morphological profiles of $A$ and $B$ layers. \cite{twesten99} Layers 
with high mobility exhibit the morphology of ripples with high peaks 
and deep valleys, while for layers with low mobility, flatter profile 
is observed. Therefore, increasing the surface mobility of either 
type of layer leads to rougher surface profile, which is more 
difficult to be smoothed out during subsequent growth.

Our analytic expressions in last subsection can also explain this 
effect. Equations (\ref{eq-lambda}) and (\ref{eq-L^*}) show that the 
value of $|\lambda|$, which determines the effective stability of the 
system, depends on parameter $\Gamma_B^*/v_B^*$. Thus, when $\Gamma_B^*$ 
is larger, deposition rate $v_B^*$ should be correspondingly increased 
in order to keep the same value of $\Gamma_B^*/v_B^*$, or to retain 
the system within the region of stability.

\subsubsection{\label{sec:asym}Interplay between mobility and global
misfit of multilayer}

In the tensile/compressive multilayer structures, one important 
parameter is the global strain of the whole multilayer film with 
respect to substrate, which varies with the relative layer thickness 
$l_B^*/l_A^*$ ($=l_B/l_A$) and relative misfit $\epsilon_B^*$
($=\epsilon_B/\epsilon_A$). A qualitative measure of the global 
strain is provided by $l_A\epsilon_A + l_B\epsilon_B$. It is 
interesting to study the stability properties of systems with 
different global misfits, in particular the \textit{asymmetry} for 
global tensile and compressive multilayers and its relationship to 
different surface mobilities of composite layers.

\begin{figure}
\resizebox{7.cm}{!}{\includegraphics{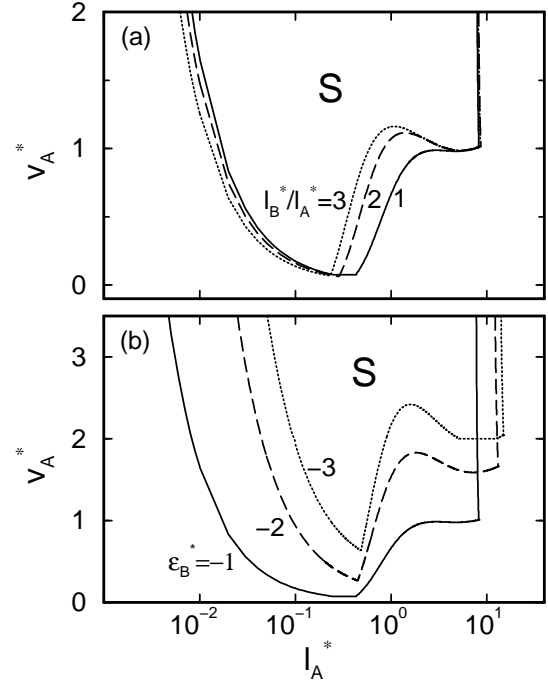}}
\caption{\label{fig-v-l-eps}Stability diagram of $l_A^*$ versus $v_A^*$, 
with parameters $\Gamma_B^*=0.01$, $v_B^*/v_A^*=1$, and $\gamma_B^*=1$.
(a) $\epsilon_B^*=-1$, with different values of $l_B^*/l_A^*$: $1$ (solid),
$2$ (dashed), and $3$ (dotted); (b) Fixed value of $l_B^*/l_A^*=1$,
but different $\epsilon_B^*=-1$ (solid), $-2$ (dashed), and $-3$ (dotted).
}
\end{figure}

\begin{figure}
\resizebox{8.cm}{!}{\includegraphics{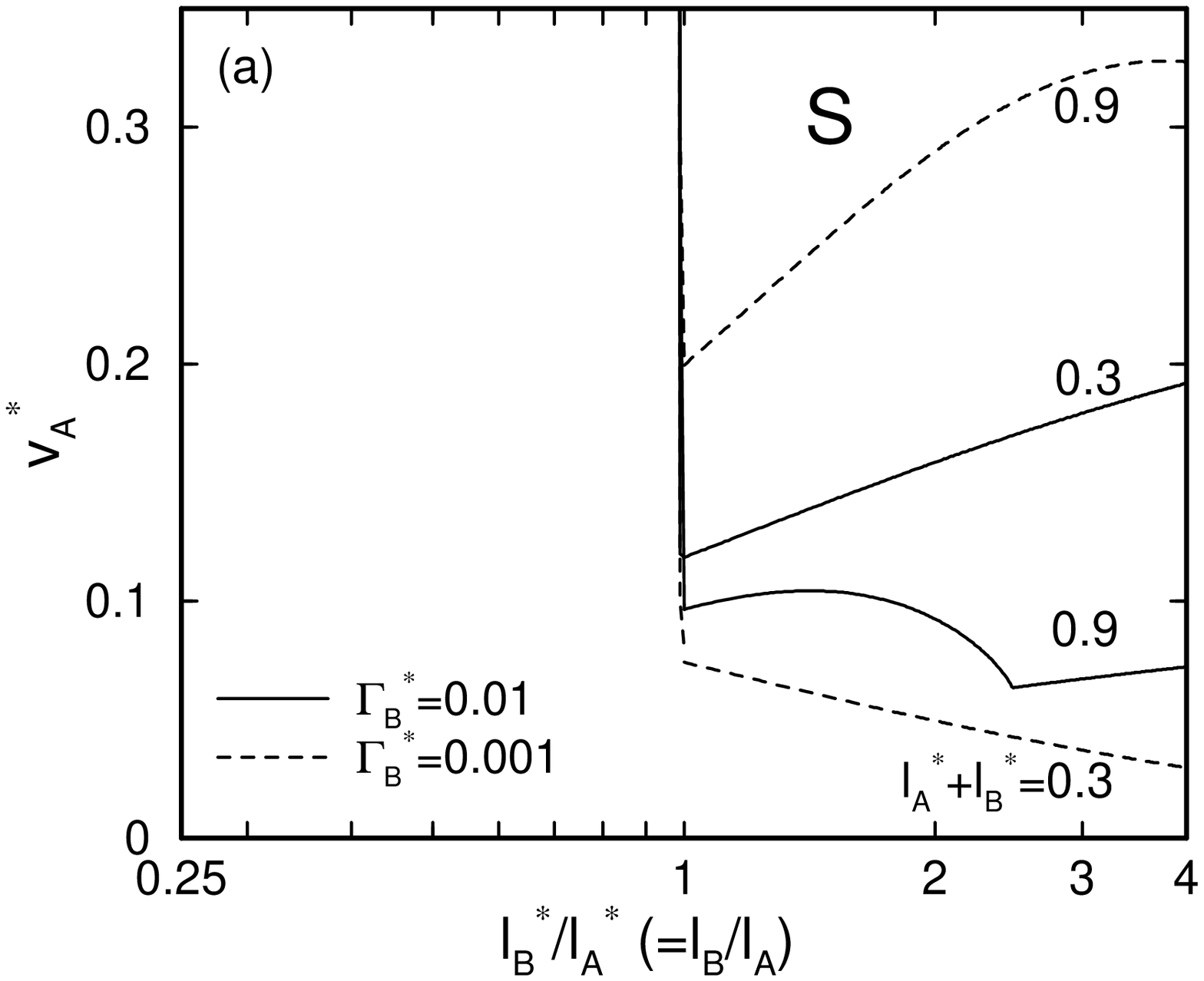}}
\vskip 0.5cm
\resizebox{8.cm}{!}{\includegraphics{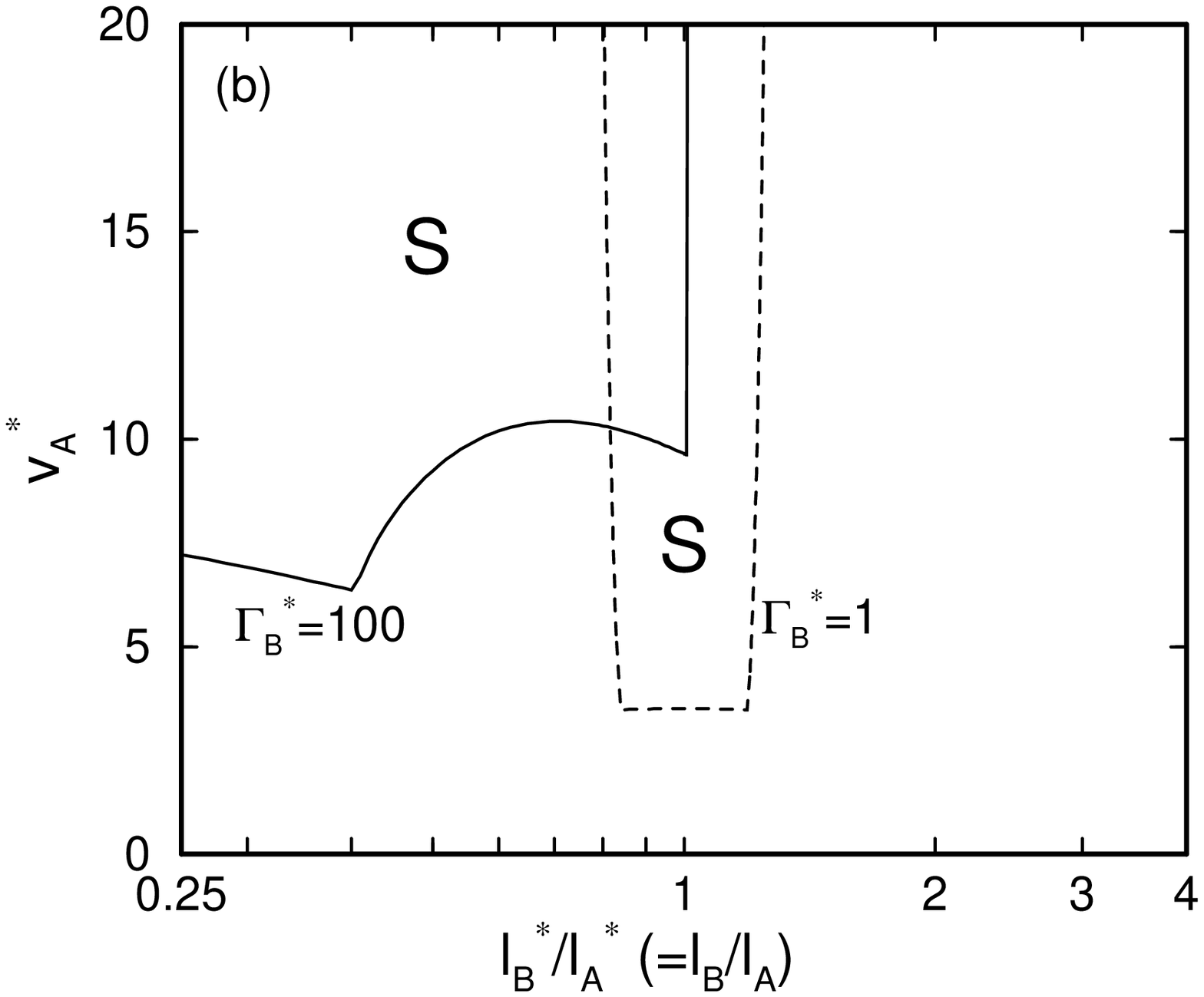}}
\caption{\label{fig-v-lBA}Stability diagrams of $l_B^*/l_A^*$ versus 
$v_A^*$, with parameters $\epsilon_B^*=-1$, $v_B^*/v_A^*=1$, and 
$\gamma_B^*=1$. (a) Relative surface mobilities $\Gamma_B^*=0.01$ 
(solid) and $0.001$ (dashed), with $l_A^*+l_B^*=0.9$ and $0.3$; 
(b) $\Gamma_B^*=100$ (solid) and $1$ (dashed), with $l_A^*+l_B^*=0.9$.
Different types of asymmetry can be seen for different values of 
$\Gamma_B^*$.}
\end{figure}

The effect of relative thickness $l_B^*/l_A^*$ (with fixed relative
misfit $\epsilon_B^*=-1$) is depicted in Figs. \ref{fig-v-l-eps} and  
\ref{fig-v-lBA}. In Fig. \ref{fig-v-l-eps} (a) we plot the stability 
diagrams of $l_A^*$ versus $v_A^*$ for different values of $l_B^*/
l_A^*$, with small relative mobility $\Gamma_B^*=0.01$. When $l_B^*/
l_A^*$ increases from $1$, the stability boundary shifts to the lower 
value of $l_A^*$; but when $l_B^*/l_A^*<1$, the system is much more 
unstable, that is, very high deposition rate $v_A^*$ (orders of 
magnitude larger than the scale shown in the figure) is needed to 
stabilize the multilayer morphology. 

Stability diagrams as a function of $l_B^*/l_A^*$ and $v_A^*$ (where 
the $A/B$ period thickness $l_A^* + l_B^*$ is held constant as in 
experiments) can illustrate this asymmetry more clearly, as shown in 
Fig. \ref{fig-v-lBA} (a) and (b). When $\Gamma_B^* <1$ ($>1$), i.e., 
$\Gamma_B <$ ($>$) $\Gamma_A$, the systems with $l_B/l_A>1$ ($<1$) 
are much more stable, corresponding to much smaller deposition rate 
$v_A^*$ ($=v_B^*$ here). When $\Gamma_B^*=1$ ($\Gamma_B=\Gamma_A$), 
Fig. \ref{fig-v-lBA} (b) shows that the stability diagram 
is much more symmetric. Also, for different relative mobilities 
$\Gamma_B^*$ [e.g., $0.01$ and $0.001$ as in Fig. \ref{fig-v-lBA} (a)], 
the system can display either a greater instability or a greater
stability as the layer period thickness $l_A^*+l_B^*$ is increased. 
This can be understood through the stability 
diagram of $l_A^*$ versus $v_A^*$ in Fig. \ref{fig-v-l}, where within 
different range of values of $l_A^*$, one can see that the stability 
can either increase or decrease with layer thickness $l_A^*$. This 
stability property with respect to period thickness is richer compared to 
the observation in InAs/AlAs experiments, where the modulation 
amplitude of the films was found to decrease with the decrease of 
period thickness. \cite{ahrenkiel98}

\begin{figure}
\resizebox{9.cm}{!}{\includegraphics{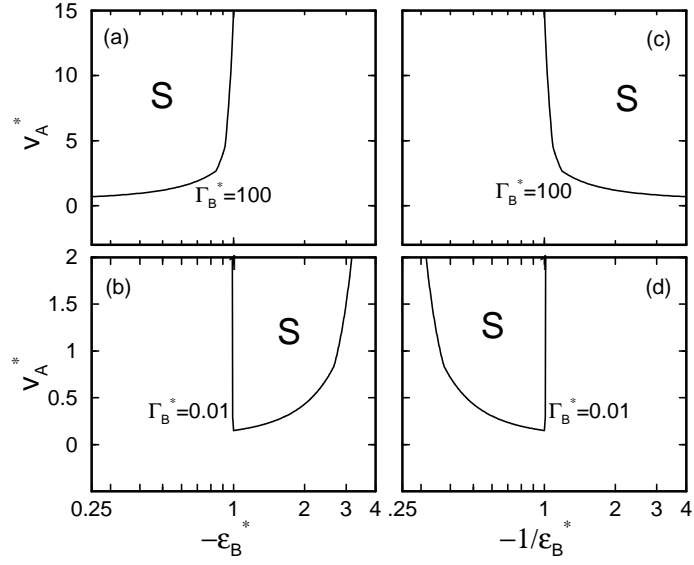}}
\caption{\label{fig-v-eps}Stability diagram of misfit $\epsilon_B^*$
versus $v_A^*$, with $l_A^*=l_B^*=0.5$, $v_B^*/v_A^*=1$, $\gamma_B^*=1$,
and different $\Gamma_B^*=100$ (a) and $0.01$ (b), showing different
types of asymmetry. Panels (c) and (d) are the replots of (a) and (b)
as $ -1 / \epsilon _B^*$ versus $v_A^*$, in order to facilitate 
comparison respectively with Fig. \ref{fig-v-lBA-bal} and its inset,
where the strain-balanced condition implies $l_B^*/l_A^*=-1/\epsilon_B^*$.
}
\end{figure}

Figures \ref{fig-v-l-eps} (b) and \ref{fig-v-eps} exhibit 
the effect of relative misfit $\epsilon_B^*$, with fixed $l_B^*/l_A^*
=1$. The asymmetric property, seen in Fig. \ref{fig-v-lBA}, also appears 
in the stability diagrams of $l_A^*$ versus $v_A^*$ for different values 
of $\epsilon_B^*$ [Fig. \ref{fig-v-l-eps} (b)]. For $\Gamma_B^*=0.01$, when 
$|\epsilon_B^*|>1$ the instability is enhanced by the increase of 
the value of misfit, with the shrinking of stability region; however, 
when $|\epsilon_B^*|<1$ the $v_A^*$ values of stability boundary are 
much larger than the scale shown here, corresponding to a very unstable 
system. We can see the asymmetry more clearly in the $\epsilon_B^*$-%
$v_A^*$ plot of Fig. \ref{fig-v-eps}, where $l_A^*=l_B^*=0.5$ and
when $\Gamma_B^* <1$ ($>1$), the systems with larger (smaller) 
misfit of layer $B$ [i.e., $|\epsilon_B^*|>1$ ($<1$)] are much more 
stable.

Consequently, these stability diagrams in Figs. \ref{fig-v-l-eps}--%
\ref{fig-v-eps} show that when the sign of the multilayer global strain 
is the same as that of the layers with \textit{smaller} surface atomic 
mobility, the system is more \textit{stable}; otherwise, when the 
layers with larger mobility have relatively larger misfit strain or 
layer thickness, the modulations on the surface of these layers are 
enhanced due to faster surface diffusion and are hard to be suppressed
by the subsequent deposition of layers with smaller mobility. Note 
that this result of stability asymmetry, which looks different for 
$\Gamma_B^*>1$ and $<1$, is in fact symmetric with respect to the 
exchange of $A$ and $B$ layers. E.g., when layer $A$ is compressive,
layer $B$ is tensile, and $\Gamma_B^*=\Gamma_B/\Gamma_A<1$, we obtain 
from above analysis that global tensile multilayer film is more
stable. If $A$ and $B$ are interchanged, layer $A'$ (of material $B$)
is tensile and layer $B'$ (of material $A$) is compressive, with
$\Gamma_{B'}^*>1$; then according to above result, the global
strain of the more stable system should be the sign of layer $A'$,
which is also tensile.

In recent experiments of InAs/AlAs short-period superlattices grown
on InP(001), \cite{norman98,ahrenkiel98,twesten99,norman02} the system 
with large tensile global strain was observed to be stable. This is 
consistent with our theoretical results, since AlAs is tensile with 
respect to substrate InP, InAs is compressive, and importantly, Al 
has much smaller surface atomic mobility, leading to the possibility 
of stabilization for global tensile multilayer (corresponding to 
thicker AlAs layers) due to the above analysis. For the global
compressive films (corresponding to thicker InAs layers), our 
calculations yield much larger instabilities, in agreement with the
most recent experimental observation \cite{norman02} that although
the multilayer structure under global compression exhibits weak
lateral modulations as a whole, the system is in fact morphologically
unstable due to a high degree of surface roughness and the reduced
strength of lateral modulations can be attributed to irregular 
surface patterns.

To further compare with the experiments of InAs/AlAs, \cite{norman98,%
ahrenkiel98,twesten99,norman02} stability diagrams of $l_B^*/l_A^*$ 
versus $v_A^*$ (Fig. \ref{fig-v-lBA}) should be used. Here layer 
$A$ corresponds to InAs and layer $B$ to AlAs. As shown in the 
figure, the stability boundary depends on the parameter values, 
in particular the relative mobility $\Gamma_B^*$ (which also depends 
on growth temperature as discussed below: when $\Gamma_B^*<1$ that 
applies here, higher temperature $T$ leads to larger $\Gamma_B^*$) 
and period thickness $l_A^*+l_B^*$. When $\Gamma_B^*$ is very small 
(e.g., $=10^{-3}$) and layer thickness $l_A^*+l_B^*$ is thin enough 
(e.g., $=0.3$), our results in Fig. \ref{fig-v-lBA} 
show that films with larger tensile global strain (corresponding to 
larger $l_B^*/l_A^*$) are more stable. However, when layer thickness 
$l_A^*+l_B^*$ increases (e.g., $=0.9$), we have opposite result for
the multilayer under global tension: higher global strain leads to 
more instability. When relative mobility is larger (e.g., $\Gamma_B^*
=10^{-2}$), different behavior can be obtained. Opposite to the case 
of $\Gamma_B^*=10^{-3}$, thinner $A/B$ layers ($l_A^*+l_B^*=0.3$ in 
Fig. \ref{fig-v-lBA}) result in more unstable global tensile film. 
Interestingly, for larger thickness $l_A^*+l_B^*=0.9$, from Fig. 
\ref{fig-v-lBA} we can get a maximum boundary value of $v_A^*$ 
(maximum instability) in the tensile region when $l_B/l_A$ is somewhat 
larger than $1$ (i.e., the whole multilayer is somewhat tensile), which 
may correspond to the experimental observation of the maximum modulation 
for slightly global tensile InAs/AlAs film. \cite{norman98,twesten99} 
Also there is a minimum boundary value $v_A^*$ for $l_A^*+l_B^*=0.9$, 
showing the maximum stability at some higher global strain.

Although our theoretical results qualitatively agrees with some
experimental observations, for the quantitative comparison more
realistic factors should be considered in our model. One important
factor is the ``wetting effect'',\cite{spencer97} especially for 
the multilayer structures like short-period superlattices where 
the thickness of each strained layer is very thin and then the 
film-substrate and layer-layer interface energies have to be 
carefully included in section \ref{sec:model}. This effect is 
absent in our calculation but applicable to the InAs/AlAs 
experimental system.

\subsubsection{\label{sec:balanced}Strain-balanced condition}

\begin{figure}
\resizebox{9.cm}{!}{\includegraphics{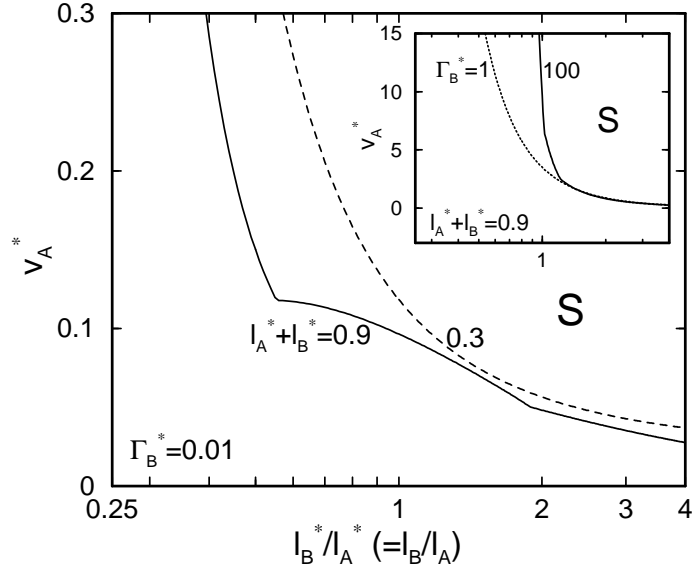}}
\caption{\label{fig-v-lBA-bal}Stability diagram of $l_B^*/l_A^*$ 
versus $v_A^*$ in the strain-balanced condition, with $v_B^*/v_A^*=1$
and $\gamma_B^*=1$. For relative mobility $\Gamma_B^*=0.01$, $l_A^*+
l_B^*$ is fixed at $0.9$ (solid) and $0.3$ (dashed). Inset: 
$\Gamma_B^*=100$ (solid) and $1$ (dotted), with $l_A^*+l_B^*=0.9$.
Compared to Figs. \ref{fig-v-lBA} and \ref{fig-v-eps}, here the 
stability property shows single type of asymmetry for different 
$\Gamma_B^*$.}
\end{figure}

In some experiments for multilayer film growth, the strains caused by
tensile and compressive layers are balanced. That is, the system
fulfills the strain-balanced condition $l_A \epsilon_A + l_B \epsilon_B 
=0$, or equivalently, $l_B^*/l_A^* = -1/\epsilon_B^*$, which makes the
global misfit of multilayer film related to substrate equal to zero.
One would expect that the corresponding stability diagrams are less 
asymmetric compared to the globally strained system studied above. 
This is seen in Fig. \ref{fig-v-lBA-bal} for diagrams of $l_B^*/l_A^*$ 
versus $v_A^*$ (with fixed $l_A^*+l_B^*$). However, the asymmetry
is still obvious, and from these stability diagrams of zero global
strain we obtain an interesting property that for all values of 
relative mobilities $\Gamma_B^*=\Gamma_B/\Gamma_A$, the multilayer 
structures are more stable, that is, can be stabilized by lower 
rescaled deposition rates $v_A^*$ and $v_B^*$, when the relative
misfit $|\epsilon_B^*| = |\epsilon_B|/|\epsilon_A| <1$ (or equivalently, 
$l_B/l_A>1$). Figure \ref{fig-v-lBA-bal} should be compared to
Fig. \ref{fig-v-eps} (c) and (d), since the later does not have a 
strain-balanced condition.

Note that this result is obtained for fixed material parameters
$\Gamma_A$, $\epsilon_A$ and $\gamma_A$ (especially for rescaled 
deposition rate $v_{A(B)}^*=v_{A(B)}/v_A^0=v_{A(B)}\gamma_A^2/(\Gamma_A 
E'^3\epsilon_A^6)$ that determines the degree of stability here), and 
one can obtain the same result with respect to fixed $B$ parameters by 
interchanging $A$ and $B$ in the calculation (see also similar diagrams 
for $\Gamma_B^*<1$ and $>1$). Thus, when we already have one layer 
material $1$ (with misfit $\epsilon_1$), to make the multilayer system 
more stable, the other layer material $2$ should be selected to have 
smaller misfit in relation to the substrate (i.e., $|\epsilon_2|
<|\epsilon_1|$), which implies a larger layer thickness $l_2>l_1$ due 
to the strain-balanced condition.

Consequently, with the condition of strain balance, the multilayer 
structure with asymmetric $A$ and $B$ layers (consisting of alternately
thin strained layers and thick layers with less misfit of opposite sign), 
are more stable compared to that with symmetric properties $\epsilon_A=
-\epsilon_B$ and $l_A=l_B$, and \textit{more the asymmetry, more the
stability.}  Also, this result is independent of the relative 
surface mobility of two types of layers. This theoretical prediction, 
which is different from the case of nonzero global strain studied above, 
should be related to different multilayer structures with different 
types of composite layers, and is yet to be verified by experiments. 

In fact, this result can lead to the well-known fact of the
heteroepitaxial growth: the system is more stable for less strained
composite layers. When we fix material $A$ and select material
$B$, the system is more stable for the material with smaller misfit
$\epsilon_B$ (and then larger thickness $l_B$) according to our result; 
then, if we fix material $B$ and select other type of layer $A$, the 
material with smaller $\epsilon_A$ corresponds to more stable system. 
The repeat of this selection procedure leads to an obvious result that 
the most stable multilayer system is the structure in which both misfits 
$\epsilon_A$ and $\epsilon_B$ are as small as possible.

\subsubsection{\label{sec:T-gamma}Role of surface tension and growth 
temperature}

\begin{figure}
\resizebox{9.cm}{!}{\includegraphics{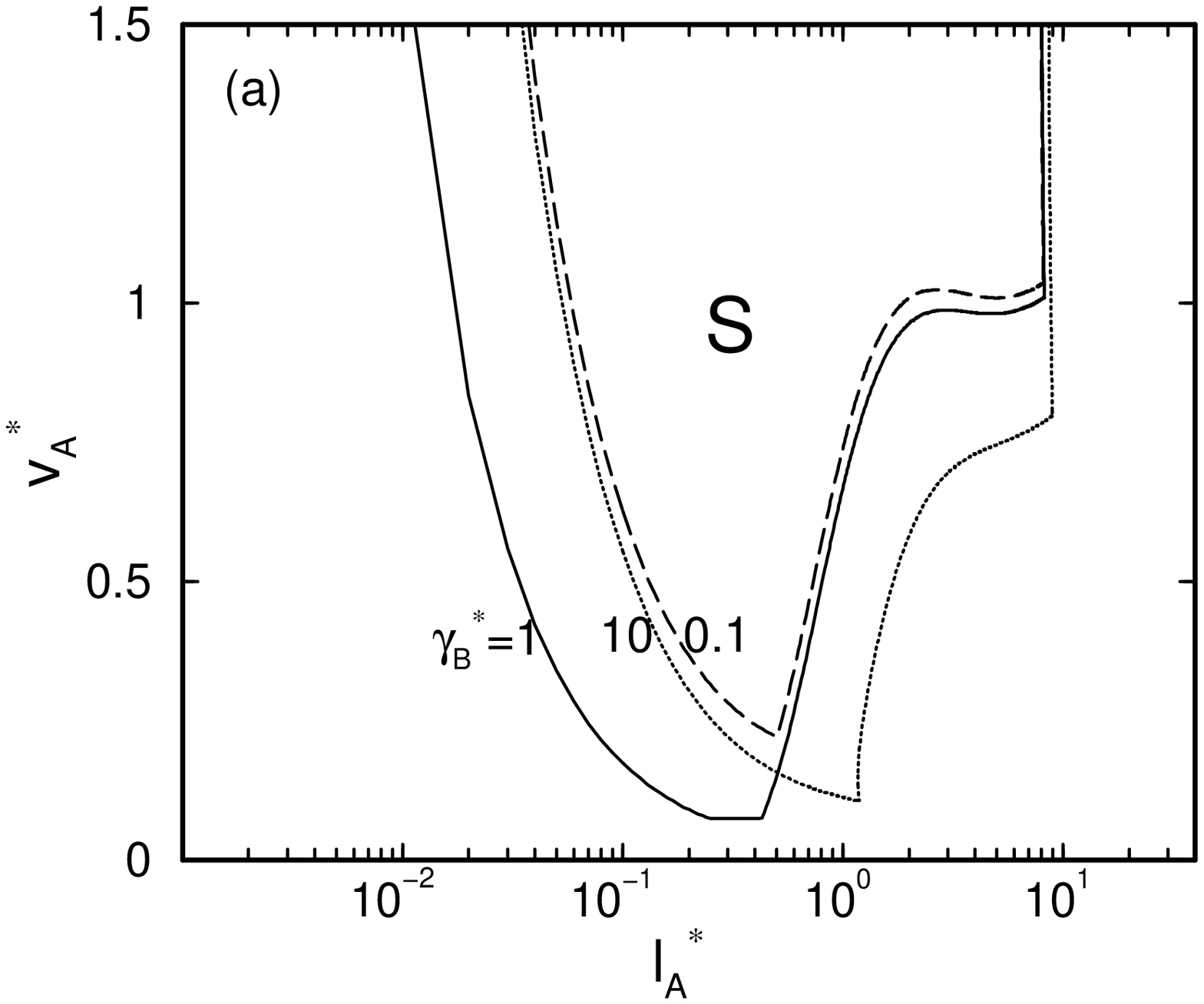}}
\vskip 0.5cm
\resizebox{9.cm}{!}{\includegraphics{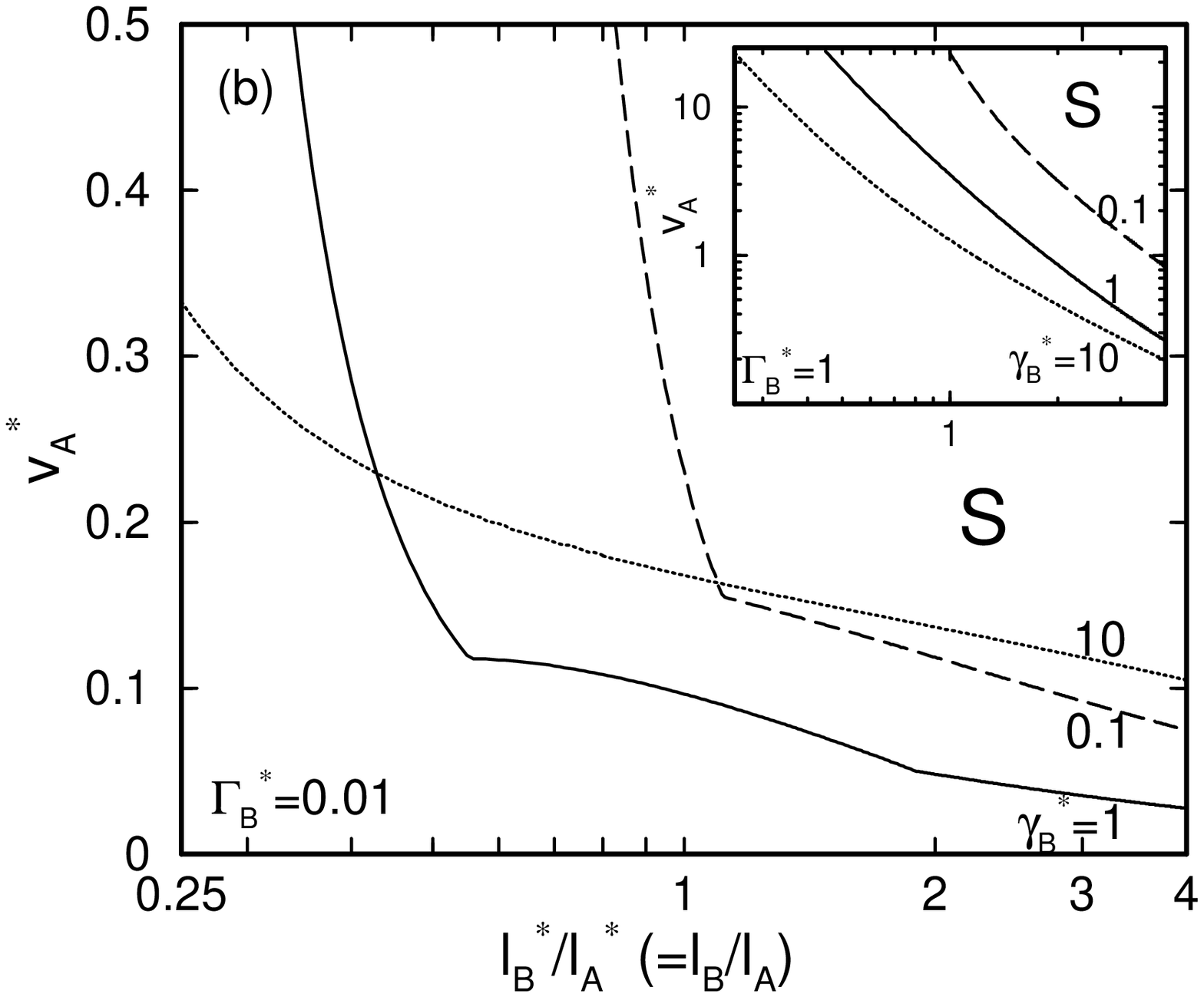}}
\caption{\label{fig-v-l-g}Stability diagrams for different relative 
surface tension $\gamma_B^*=0.1$ (dashed), $1$ (solid), and $10$
(dotted), with $v_B^*/v_A^*=1$. In (a), $v_A^*$ is plotted as a
function of $l_A^*$, with $\Gamma_B^*=0.01$,  $\epsilon_B^*=-1$, 
$l_B^*/l_A^*=1$, and $v_B^*/v_A^*=1$; while in (b), the system
is in the strain-balanced condition, and we use different $\Gamma_B^*$
[$0.01$ and $1$ (inset)] in the stability diagram of $l_B^*/l_A^*$ 
versus $v_A^*$, with fixed $l_A^*+l_B^*=0.9$.}
\end{figure}

Our results for stability properties are not very sensitively 
dependent on relative surface tension $\gamma_B^*=\gamma_B/\gamma_A$ 
(One of the major reasons may be that wetting effect \cite{spencer97} 
is not considered here). See diagrams of $l_A^*$ versus $v_A^*$ in 
Fig. \ref{fig-v-l-g} (a), with $\Gamma_B^*=0.01$ and different 
$\gamma_B^*=0.1$, $1$, and $10$. For relative tension $\gamma_B^*<1$, 
the stable region is smaller; but the stability diagrams do not 
change much if we compare the result of $\gamma_B^*=10$ to that 
of $0.1$ (with the difference of two orders of magnitude for 
$\gamma_B^*$).

We also calculate the diagrams as a function of $l_B^*/l_A^*$ and 
$v_A^*$ in the strain-balanced condition, for different relative 
surface tension $\gamma_B^*$, as shown in Fig. \ref{fig-v-l-g} 
(b). The results depend on the relative mobility $\Gamma_B^*$ and 
period thickness $l_A^*+l_B^*$. For $l_A^*+l_B^*=0.9$ and $\Gamma_B^*
=1$, the inset of Fig. \ref{fig-v-l-g} (b) shows that the stability 
of the system increases with the increase of $\gamma_B^*$, while 
for smaller $\Gamma_B^*=0.01$, the results are more complicated: 
for large $l_B/l_A$, $\gamma_B^*=1$ (i.e., $\gamma_A=\gamma_B$) 
corresponds to the most stable case, but for small $l_B/l_A$, larger 
$\gamma_B^*$ favours the stability.

Growth temperature is an important factor in real experiments. 
Our results in Sec. \ref{sec:A/B-formul} show that the stability 
properties are affected by growth temperature $T$ via mobilities 
$\Gamma_A$ and $\Gamma_B$, since the rescaled parameter 
$\Gamma_B^*$ changes exponentially with $T$: $\Gamma_B^* \propto 
\exp[-(E_B-E_A)/k_BT]$ with $E_A$ and $E_B$ the activation energies 
of $A$ and $B$ materials. If $E_B \geq E_A$, as temperature $T$ 
increases, value of $\Gamma_B^*$ increases or remains unchanged;
while for $E_B < E_A$, $\Gamma_B^*$ decreases with the increment
of $T$. Also, the dimensionless deposition rates $v_A^*$ and $v_B^*$ 
decrease with temperature due to $v_{A(B)}^*=v_{A(B)}\gamma_A^2/
(\Gamma_A E'^3 \epsilon_A^6)$ [see Eqs. (\ref{eq-v_A^0}) and 
(\ref{eq-trans})], $\Gamma_A \propto T^{-1}\exp(-E_A/k_BT)$, and
then $v_{A(B)}^* \propto T\exp(E_A/k_BT)$. Thus, when temperature 
$T$ increases, for the case of $E_B \geq E_A$ (i.e., $\Gamma_B^* 
\leq 1$) larger (or fixed) value of $\Gamma_B^*$ and smaller 
deposition rate $v_{A(B)}^*$ render the system more unstable, 
according to the stability diagrams in e.g., Fig. \ref{fig-v-l}. 
Similar result can be obtained for $E_B < E_A$ (i.e., $\Gamma_B^*>1$). 
As discussed above, when $T$ increases, value of $v_{A(B)}^*$ gets 
smaller, leading to more unstable system, but $\Gamma_B^*$ also 
decreases, favoring the stability of the film. Note that due
to $0<E_A-E_B<E_A$, the reduction of $\Gamma_B^*$ is much slower 
than that of $v_{A(B)}^*$ as $T$ is increased. Hence, we still 
have the result that higher growth temperature enhances the 
instability of this multilayer structure, as expected.

\subsection{\label{sec:A/B-disc}Discussion}

For the alternating tensile/compressive multilayer, when a layer $i$, 
which is tensile (compressive) with respect to the substrate, is 
grown with the presence of undulating surface, local strain energy 
concentration is created at the surface troughs with more stress 
relaxation at the peaks. When the next compressive (tensile) layer
$i+1$ containing larger (smaller) atoms is deposited on it, the 
difference of strain energy density along the underlying surface 
drives the deposited atoms to the trough regions. \cite{shilkrot00} 
This corresponds to the coupling of strain fields for two successive 
upper layers, while in reality the strain coupling is more complicated 
due to the influence of lower buried layers, which has been shown to
decay exponentially with the distance from the growing surface [see 
Eq. (\ref{eq-h_k-q})]. Nevertheless, some of our results 
in Sec. \ref{sec:A/B-results} can be explained by this simple 
two-layer coupling picture. When the layer thickness is too small,
that is, the deposited material is not enough to smooth out the
underlying troughs, the subsequent deposited layer $i+2$ will deepen
the corrugations since it is of the same material as layer $i$. Thus,
the surface perturbation is amplified by subsequent depositions of
different layers, rendering the system unstable. On the other hand,
when the layer thickness is too large, we have inverse morphology,
that is, peak (valley) forms and continues to grow above the valley 
(peak) of underlying layer $i$, and then the surface morphological
modulations can not decay with the layer stacking. Thus, for 
intermediate range of layer thickness, the multilayer is more stable,
i.e., lower deposition rate is needed to reduce and suppress the
surface undulations during each layer growth. Here, surface mobility
plays an important role. Smaller mobility leads to slower stress-driven
surface diffusion process, and then the surface morphology has less
dependence on the corrugations of underlying layer. Consequently,
the stability results in Fig. \ref{fig-v-l} can be reproduced, that 
is, the system is more unstable for too small or too large a layer 
thickness, higher relative mobility, and smaller deposition rate. 

More interesting results are found for the multilayer films with
global misfit with respect to the substrate. When the layers with 
smaller surface mobility have larger thickness or larger magnitude 
of misfit, that is, the sign of the multilayer global strain is
the same as that of this type of layers, the system is more stable.
This can also be seen in the above approximate picture of two-layer 
coupling. For example, for the growth of layer with small mobility, 
the surface evolution does not depend much on the undulations of the 
layer underneath, and then with large thickness the surface profile 
is flatter than that of underlying layer (as observed in experiments); 
when the next layer with larger mobility is deposited, the surface 
corrugation develops but is limited due to the smaller layer 
thickness. Therefore, the modulations are easier to suppress
during the layer stacking procedure. This asymmetric property is
consistent with the recent experimental results of InAs/AlAs on
InP(001), where the films under high global tension have been found 
to be stable \cite{norman98,ahrenkiel98,twesten99,norman02} while
those under global compression are much more unstable. \cite{norman02}
Note that the formation of lateral modulations observed in the whole 
multilayer film is due to the regular surface modulated pattern; for 
irregular surface pattern weak lateral modulations are observed, but 
the surface could be very rough, corresponding to unstable system.
\cite{norman02}

The other property we obtain for the tensile/compressive multilayer
is related to the films with strain-balanced condition, i.e., zero
global strain. We find that with this constraint, the multilayer 
structure is more stable if one type of strained layers is thicker 
and then has less misfit of opposite sign compared to the other 
type. To see this in experiments, one should compare multilayer 
films composed of different materials (such as $A/B$ and $A/B'$ 
structures both with zero global strain, but $l_B \neq l_B'$ and 
$\epsilon_B \neq \epsilon_{B'}$). As discussed in Sec. 
\ref{sec:balanced}, this asymmetry property in fact leads to 
the well-known result that smaller misfit of each composite layer 
results in more stable system.

\section{\label{sec:spacer}Strained/spacer multilayer structures}

The other kind of multilayer that we are interested in is the
structure with strained/spacer stacked layers. In this system, the
strained layers (either tensile or compressive with respect to the
substrate) and spacer layers are alternately deposited on a substrate
of material $B$, such as InAs/GaAs(001), \cite{xie95,gonzalez00,%
nakata97} Ge/Si(001), \cite{schmidt99,lethanh99,lethanh00} and 
SiGe/Si(001). \cite{tersoff96,teichert96,lafontaine98} The strained 
layers are odd-numbered ($i=1$, $3$, $...$), with nonzero misfit 
strains $\epsilon_i \neq 0$, while the even-numbered spacer layers 
($j=2$, $4$, $...$) are of substrate material $B$, with zero misfit 
$\epsilon_j=\epsilon_B = 0$. Here, we consider the case in which 
all the spacer layers are thick enough, such that the interface 
between a spacer layer $j$ and a strained layer $(j+1)$ is atomically 
flat ($\hat{h}_j = 0$ for even $j$) since the growth front of 
previously deposited strained layer is smoothed out. [If in the 
solution (\ref{eq-solu}), one sets $\epsilon_k=\epsilon_B = 0$,
then $\sigma_k$ is negative and for thick enough spacer layers,
Eq. (\ref{eq-solu}) leads to a flat top surface of the spacer layer.
This is also seen in many experiments.]

The experimental interest for this multilayer structure is different 
from that of tensile/compressive multilayer due to the presence of 
spacer layers. This system is morphologically unstable (as verified
below), and 3D islands develop for each strained layer with the 
self-organized formation of multisheet arrays of islands along the
growth direction. It has been observed that both size and regularity 
of islands increase with the increment of deposition layers, and
then saturate for large enough layer numbers. In the linear analysis 
presented here, this property of the island size can be explained 
through the study of kinetic critical thickness for the onset of 
3D instability, with details shown in Sec. \ref{sec:spacer-B}.

\subsection{\label{sec:spacer-A}Elastic analysis and morphological
evolution}

The recursive procedure for elastic analysis presented in Sec. 
\ref{sec:model} can be applied to the strained/spacer multilayer
structures. The corresponding results are simpler due to the
zero misfit of spacer layers and the flat strained/spacer 
interfaces (i.e., $\epsilon_j=0$, $\hat{h}_j = 0$ for even 
$j$), and are different for the odd-numbered (strained) and 
even-numbered (spacer) layers. Consider a $k$-layer system with
top layer $k$ strained with respect to the substrate (i.e., with
odd number $k$). Thus, the interfaces between odd- and even-%
numbered layers (that is, $i/(i-1)$ interfaces with odd $i$ and
$3 \leq i \leq k$) in Fig. \ref{fig-sche} (a) are planar. The
effective boundary conditions at the top surface $z=\bar{h}_k$
and the interfaces $z=\bar{h}_i$ ($0 \leq i \leq k-2$) below 
layer $(k-1)$ are the same as Eqs. (\ref{eq-bound1-n}) and 
(\ref{eq-bound3-n}), while at planar $k/(k-1)$ interface the 
special boundary condition (\ref{eq-bound2-n}) should be revised
as ($\alpha=x$, $y$, $z$)
\begin{equation}
\hat{\sigma}_{\alpha z,k}|_k = \hat{\sigma}^n_{\alpha z,k-1}|_k
\qquad {\rm and} \qquad 
\hat{u}_{\alpha,k}|_k = \hat{u}_{\alpha,k-1}|_{k-1}
+ \hat{u}^n_{\alpha,k-1}|_k.
\label{eq-bound2'-n}
\end{equation}
Similar to the discussion in Sec. \ref{sec:model}, we obtain an
elastically effective system composed of an undulating top free 
surface and a \textit{planar} interface between top single layer
$k$ and an effective semi-infinite elastic medium of field 
$u_{\alpha,k-1}^n|_k$ with boundary condition (\ref{eq-bound2'-n}),
to first order of perturbations and with the assumption of identical
elastic constants. To determine the elastic fields recursively, we 
need to consider the system with smaller number of layers. For a
$m$-layer system with $1 < m < k$, the above results can be directly
applied with the replacement of $k$ by $m$ if layer number $m$ is
odd. However, for even number $m$, the results are somewhat different 
due to the flat top surface of even-numbered spacer layer $m$. The
corresponding effective system is in fact composed of top layer $m$,
with a \textit{planar} free surface, and underlying effective 
semi-infinite medium of elastic field $u_{\alpha,m-1}^n|_m$, with
a \textit{nonplanar} buried interface. The boundary conditions 
are different from those of Eqs. (\ref{eq-bound1-n}) and 
(\ref{eq-bound2'-n}): At planar top free surface $z=\bar{h}_m$,
\begin{equation}
\hat{\sigma}_{\alpha z,m}|_m = 0,
\label{eq-bound1''-n}
\end{equation}
while at rough interface $z=\bar{h}_{m-1}$,
\begin{subequations}
\label{eq-bound2''-n}
\begin{equation}
\hat{\sigma}_{\alpha z,m}|_m = \hat{\sigma}_{\alpha z,m-1}^n|_m,
\label{eq-bound2a''-n}
\end{equation}
\begin{eqnarray}
&\hat{u}_{\beta,m}|_m = \hat{u}_{\beta,m-1}|_{m-1}
+ \hat{u}^n_{\beta,m-1}|_m \quad {\rm and}& \nonumber\\
& -\bar{u}_{m-1} \hat{h}_{m-1} +\hat{u}_{z,m}|_m
=\hat{u}_{z,m-1}|_{m-1}+ \hat{u}^n_{z,m-1}|_m.&
\label{eq-bound2b''-n}
\end{eqnarray}
\end{subequations}
(Here $\alpha=x$, $y$, $z$ and $\beta=x$, $y$.)

According to the procedure in sections \ref{sec:model} and 
\ref{sec:general}, we can determine the elastic energy density
and then derive the evolution equation for surface morphological
perturbation of top strained layer $k$:
\begin{eqnarray}
\partial \hat{h}_k(q,t) / \partial t &=& \sigma_k \hat{h}_k(q,t) +
E'\Gamma_k \epsilon_k q^3 e^{-q v_k t} \nonumber\\
&\times& \left [ \sum\limits_{j=1}^{\frac{k-1}{2}}
e^{-q\sum\limits_{i=2j}^{k-1} l_i} \epsilon_{2j-1}\hat{h}_{2j-1}(q) 
- e^{-q\sum\limits_{i=1}^{k-1} l_i} \epsilon_1 \hat{\zeta}(q) 
\right ],
\label{eq-h-hat}
\end{eqnarray}
which in fact can be obtained from Eq. (\ref{eq-h_k-q}) of general 
case by considering $\epsilon_j=0$ and $\hat{h}_j = 0$ for spacer 
layer $j$. For simplicity, here we assume the planar film-substrate 
interface, i.e., $\hat{\zeta}=0$, and the initial perturbation on 
the top surfaces of substrate and of all the spacer layers, from which 
the strained layers start growing, to be $\hat{h}^0(q)$, which can 
arise from the background roughness due to the noise of deposition 
flux even if the starting surface is perfectly flat. Then the 
solution of evolution equation (\ref{eq-h-hat}) is expressed as
\begin{equation}
\hat{h}_k(q,t) = A_k(q,t) \hat{h}^0(q),
\label{eq-h-A}
\end{equation}
where $A_k(q,t)$ is the morphological perturbation amplitude for top
surface of strained layer $k$. For $k=1$, $A_1(q,t)=\exp(\sigma_1 t)$ 
according to Eq. (\ref{eq-h1-t}) and the result of single-layer film 
growth; for $k=3$, we have
\begin{equation}
A_3(q,t) = e^{\sigma_3 t} + \frac{\Gamma_3 E'\epsilon_1\epsilon_3 q^3}
{qv_3+\sigma_3}~ e^{\sigma_1 l_1/v_1 - q l_2}
\left ( e^{\sigma_3 t} - e^{-q v_3 t} \right );
\label{eq-A3}
\end{equation}
and for $k\geq 5$,
\begin{eqnarray}
A_k(q,t) &=& L_{k,1}(q,t) + L_{k,2}(q,t) \Bigglb \{
\bar{L}_{k-2,1}(q) \nonumber\\
&+& \sum\limits_{i=2}^{\frac{k-5}{2}} \left [ \prod\limits_{j=i+1}
^{\frac{k-3}{2}} \bar{L}_{2j+1,2}(q) \right ] \bar{L}_{2i+1,1}(q)
+\left [ \prod\limits_{j=2}^{\frac{k-3}{2}} \bar{L}_{2j+1,2}(q) 
\right ] \bar{A}_3(q) \Biggrb \},
\label{eq-A}
\end{eqnarray}
where
\begin{eqnarray}
L_{k,1}(q,t) &=& e^{\sigma_k t} - 
\frac{\Gamma_k \epsilon_k}{\Gamma_{k-2} \epsilon_{k-2}}~ 
\frac{qv_{k-2}+\sigma_{k-2}}{qv_k+\sigma_k}~
\frac{e^{\sigma_k t} - e^{-qv_k t}}{e^{\sigma_{k-2}l_{k-2}/v_{k-2}}
-e^{-ql_{k-2}}} ~ e^{\sigma_{k-2}l_{k-2}/v_{k-2}-q(l_{k-1}+l_{k-2})},
\nonumber\\
L_{k,2}(q,t) &=& \frac{\Gamma_k \epsilon_k}{\Gamma_{k-2} 
\epsilon_{k-2}}~ \frac{e^{-ql_{k-1}}}{qv_k+\sigma_k}~
\frac{e^{\sigma_k t} - e^{-qv_k t}}{e^{\sigma_{k-2}l_{k-2}/v_{k-2}}
-e^{-ql_{k-2}}} \nonumber\\
&\times& \left [ \left ( qv_{k-2} - \Gamma_{k-2}\gamma_{k-2} q^4 
\right ) e^{-ql_{k-2}} + \Gamma_{k-2} E' \epsilon_{k-2}^2 q^3
e^{\sigma_{k-2}l_{k-2}/v_{k-2}} \right ],
\label{eq-L}
\end{eqnarray}
and $\bar{A}_3(q)$, $\bar{L}_{2j+1,1}(q)$, and $\bar{L}_{2j+1,2}(q)$
[$2 \leq j \leq (k-3)/2$] are obtained by replacing $k$ with $2j+1$
and $t$ with $l_{2j+1}/v_{2j+1}$ in $A_3(q,t)$, $L_{k,1}(q,t)$, and
$L_{k,2}(q,t)$, respectively. Formula (\ref{eq-L}) shows that when 
the underlying spacer layer is very thick, i.e., $l_{k-1} \gg 1$, 
we have $L_{k,1}(q,t) \rightarrow \exp(\sigma_k t)$ and
$L_{k,2}(q,t) \rightarrow 0$. Then the single-layer solution
similar to Eq. (\ref{eq-h1-t}): $\hat{h}_k(q,t)=\exp(\sigma_k t) 
\hat{h}^0(q)$ is recovered for this multilayer structure with
any value of the layer number $k$, as expected.

Consequently, we have derived the explicit solution [Eqs. 
(\ref{eq-h-A})--(\ref{eq-L})] for the surface morphological 
perturbation, in terms of material and growth parameters for all 
the deposited layers ($1 \leq i \leq k$): $\Gamma_i$, $\epsilon_i$, 
$\gamma_i$, $l_i$, $v_i$, and the elastic constant $E'$. Also, the 
above expressions can be directly used to determine the early 
evolution and stability properties of stacked strained/spacer 
structures, for which the composite materials and growth conditions 
of different layers could be different (non-periodic structure) or 
the same (periodic structure).

\subsection{\label{sec:spacer-B}Results: kinetic critical thickness}

Here we apply the general results derived above to periodic 
strained/spacer structure studied in many experiments, 
\cite{tersoff96,teichert96,xie95,gonzalez00,nakata97,schmidt99,%
lethanh99,lethanh00,lafontaine98} for which, all the strained layers
(odd-numbered) consist of the same material $A$, with the same 
deposition rate $v_A$ and average thickness $l_A$, and the thickness 
of each spacer layer (even-numbered) is fixed as $l_B$. Thus, from 
Eq. (\ref{eq-A}) the perturbation amplitude $A_k$ (for odd $k$ and
$k\geq 5$) is simplified as
\begin{equation}
A_k(q,t) = L_1(q,t) + L_2(q,t) \left [ \bar{L}_1(q) \frac{1-
\bar{L}_2(q)^{\frac{k-5}{2}}}{1-\bar{L}_2(q)} 
+ \bar{L}_2(q)^{\frac{k-5}{2}} \bar{A}_3(q) \right ],
\label{eq-A_k}
\end{equation}
where $L_{1(2)}(q,t)$ corresponds to $L_{k,1(2)}(q,t)$ in formula
(\ref{eq-L}) in which one has set $\sigma_i=\sigma_A$, $\Gamma_i=
\Gamma_A$, $\epsilon_i=\epsilon_A$, $\gamma_i=\gamma_A$, $l_i=l_A$, 
and $v_i=v_A$ ($i=k$, $k-2$), and $\bar{L}_{1(2)}(q)=L_{1(2)} 
(q,t=l_A/v_A)$, so that all of them are independent of layer 
number. For smaller $k$, the amplitudes are $A_1(q,t) = 
\exp(\sigma_A t)$ for single-layer film, and
\begin{equation}
A_3(q,t) = e^{\sigma_A t} + \frac{\Gamma_A E' \epsilon_A^2 q^3}
{qv_A+\sigma_A}~ e^{\frac{\sigma_A l_A}{v_A} - q l_B}
\left ( e^{\sigma_A t} - e^{-q v_A t} \right )
\label{eq-A_3}
\end{equation}
as obtained directly from Eq. (\ref{eq-A3}).

\begin{figure}
\resizebox{9.cm}{!}{\includegraphics{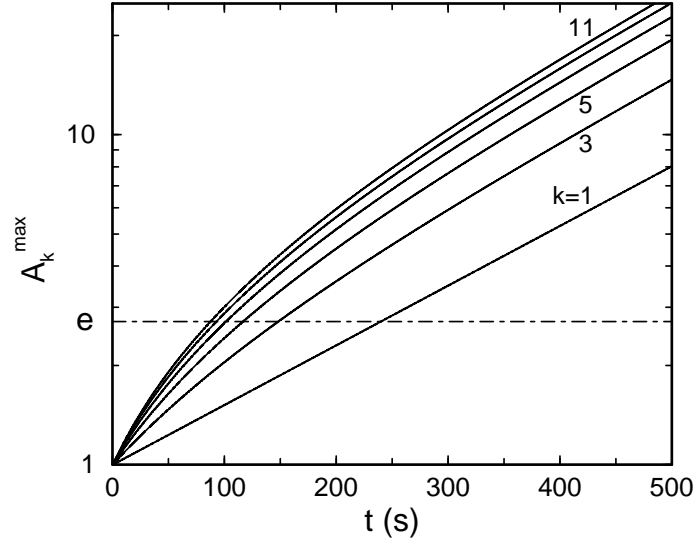}}
\caption{\label{fig-Akt}Maximum perturbation amplitude $A_k^{\rm max}$ 
versus time $t$, for successive strained layer numbers $k=1,3,...,11$
(from bottom to top). Note the semilog plot. The material parameters 
used are analogous to those of Ge/Si(001) growth system, with $l_A=4$ ML, 
$l_B=9$ nm, $v_A=1$ ML/min, and $T=550$ $^{\circ}$C.}
\end{figure}

>From Eqs. (\ref{eq-A_k}) and (\ref{eq-A_3}) we find that 
when the growth time is large enough, i.e., $t \gg 1$, the
perturbation amplitude $A_k(q,t)$ is proportional to 
$\exp(\sigma_A t)$, with positive perturbation growth rate
for small wavenumber $q$ as in the case of single-layer film.
In Fig. \ref{fig-Akt}, we show the maximum value of perturbation
amplitude (with respect to wavenumber $q$) $A_k^{\rm max}(t)$ 
as a function of growth time $t$ for different layer numbers, 
and note the positive growth rate.
Thus, similar to the tensile/compressive structure discussed
in Sec. \ref{sec:A/B}, essentially this strained/spacer multilayer
system is morphologically unstable. However, different from the
tensile/compressive films, the system here cannot be effectively
stabilized (with the decay of initial morphological perturbation) 
in certain range of growth parameters. This is due to the fact 
that the surface perturbation amplitude always increases with 
layer number $k$, i.e., $A_k(q,t) > A_{k-2}(q,t) > ...> A_3(q,t) 
> A_1(q,t) >0$ for certain wavenumber $q$ and growth time $t$ 
($=l_A/v_A$), which can be proved analytically from Eqs. 
(\ref{eq-A_k}) and (\ref{eq-A_3}) (see also numerical result of
maximum perturbation amplitude shown in Fig. \ref{fig-Akt}),
and can be attributed to the coupling of interface perturbations
of all the buried layers exhibited in Eq. (\ref{eq-h-hat}). 
Also, if the spacer layer thickness $l_B$ is large enough that 
$\bar{L}_2(q)<1$ for any mode $q$, we find the asymptotic
expression for amplitude $A_k$, as $k \rightarrow \infty$, as
\begin{equation}
A_{k}(q,t) \rightarrow L_1(q,t) + L_2(q,t) \bar{L}_1(q)
/[1-\bar{L}_2(q)].
\label{eq-Ak-infty}
\end{equation}

Nevertheless, during the growth process the surface instability of top 
strained layer can be \textit{kinetically} stabilized up to a critical
thickness, as in the single-layer film growth. \cite{spencer91} For
growing layer below this kinetic critical thickness, that is, at early 
time of layer deposition, the surface perturbation does not develop 
fast enough compared to the deposition of materials flux, and then 
is not observable. The perturbation becomes apparent relative to the 
background disturbance only when the layer is sufficiently thick and 
then the initial undulation $\hat{h}^0$ for the growth of top strained 
layer is substantially amplified. As in the study of single-layer film, 
\cite{spencer00} we use $\ln(\hat{h}/\hat{h}^0)\sim 1$ as the 
amplification limit, or equivalently, we use
\begin{equation}
\ln \left [ A_k^{\rm max} (t=t_c) \right ] = 1
\label{eq-Amax}
\end{equation}
with $t_c=h_c|_k/v_A$, as a measure of the kinetic critical thickness
$h_c|_k$ for a $k$-layer system (see Fig. \ref{fig-Akt}). In Eq. 
(\ref{eq-Amax}), we have assumed that the surface morphological 
property is dominated by the mode of maximum perturbation amplitude 
($A_k^{\rm max}$), that is, by the wavelength of the strongest 
instability which, in single-layer case, may correspond to the 
wavelength of the surface roughness pattern observed in real 
experiments. \cite{lagally00,tromp00}

\begin{figure}
\resizebox{9.cm}{!}{\includegraphics{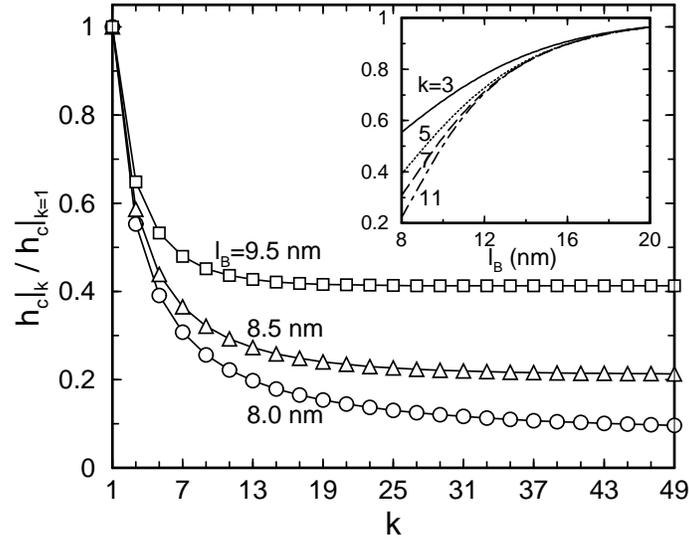}}
\caption{\label{fig-hc-k}Kinetic critical thickness $h_c|_k$ (rescaled 
by that of single layer $h_c|_{k=1}$) as a function of the layer number 
$k$, with the Ge/Si(001) growth parameters: $l_A=4$ ML, $v_A=1$ ML/min, 
$T=550$ $^{\circ}$C, and different spacer layer thickness $l_B=9.5$ nm 
(square), $8.5$ nm (triangle), and $8$ nm (circle). 
Inset: $h_c|_k/h_c|_{k=1}$ as a function of $l_B$, for $k=3$, $5$, 
$7$, and $11$.}
\end{figure}

The rescaled kinetic critical thickness $h_c|_k/h_c|_{k=1}$ (with
$h_c|_{k=1}$ the result for single-layer growth) is plotted as a
function of layer number $k$ in Fig. \ref{fig-hc-k}, according to
Eqs. (\ref{eq-A_k}) and (\ref{eq-Amax}). We choose the material
parameters of Ge/Si(001) multilayer growth, \cite{lethanh99,%
lethanh00} with strained layer thickness $l_A=4$ ML, deposition 
rate $v_A=1$ ML/min, temperature $T=550$ $^{\circ}$C, and different 
spacer layer thickness $l_B=8$ nm, $8.5$ nm, and $9.5$ nm. (Typical 
Ge layers are rather thin and usually expressed in units of 
monolayers, while the Si spacer layers are relatively thick and 
appropriately expressed in nanometers, as in experiments. 
\cite{schmidt99,lethanh99,lethanh00} We use the conversion $1$ ML 
$=0.1457$ nm for Ge monolayers \cite{lethanh99} in our calculations.)
For surface mobility $\Gamma_k$, the definitive measurement of its 
absolute value over a range of temperature is lacking, and we 
determine it from the expression of single-layer result: $h_c|_{k=1}
=256v_A\gamma_A^3/(27\Gamma_A\epsilon_A^8 E'^4)$, as well as the 
observed value of $h_c|_{k=1}$, corresponding to the experimentally 
measured 2D to 3D transition thickness, in Ge/Si(001) growth. As 
illustrated in Fig. \ref{fig-hc-k}, the kinetic critical thickness 
$h_c|_k$ first decreases monotonically and rapidly as the number $k$ 
of deposited layers increases, and then slowly until it converges to 
a saturation value at large enough number $k$. This convergence effect 
is due to the asymptotic result of perturbation amplitude $A_k$ [see 
Eq. (\ref{eq-Ak-infty})] when the spacer layer is sufficiently thick
and then $\bar{L}_2(q)<1$. Also, from the comparison of different 
symbols (corresponding to different spacer layer thicknesses $l_B$) 
in Fig. \ref{fig-hc-k}, we find that $h_c|_k$ is reduced more 
rapidly for thinner spacer layer, approaching a lower saturation 
value at larger number $k$. This can be attributed to the stronger 
coupling between top surface and the underlying layers and interfaces 
for smaller $l_B$, as seen in Eq. (\ref{eq-h-hat}).

These results shown in Fig. \ref{fig-hc-k}, both for the behavior 
of $h_c|_k$ versus $k$ and for the effect of spacer thickness $l_B$, 
are in good agreement with the recent experimental observations 
\cite{nakata97,schmidt99,lethanh99,lethanh00} of critical thickness 
for 2D-3D growth transition. The properties of kinetic critical 
thickness presented here can explain some experimental results for
islands (quantum dots) growth. When the thickness of growing film
exceeds $h_c|_k$, the morphological instability occurs, exhibiting
as surface ripples or corrugations which have been observed as the 
precursor to the coherent islands formation. \cite{lagally00,tromp00}
Thus, smaller $h_c|_k$ not only corresponds to the earlier occurrence 
of islands during growth, but also results in larger island size 
since it implies more time for the development of 3D islands if the
nominal thickness of top strained layer is fixed. Consequently, the
results in Fig. \ref{fig-hc-k} correspond to the experimental 
phenomenon of the increase and then saturation of islands size in
upper strained layers. \cite{nakata97,lethanh99,lethanh00}

\begin{figure}
\resizebox{9.cm}{!}{\includegraphics{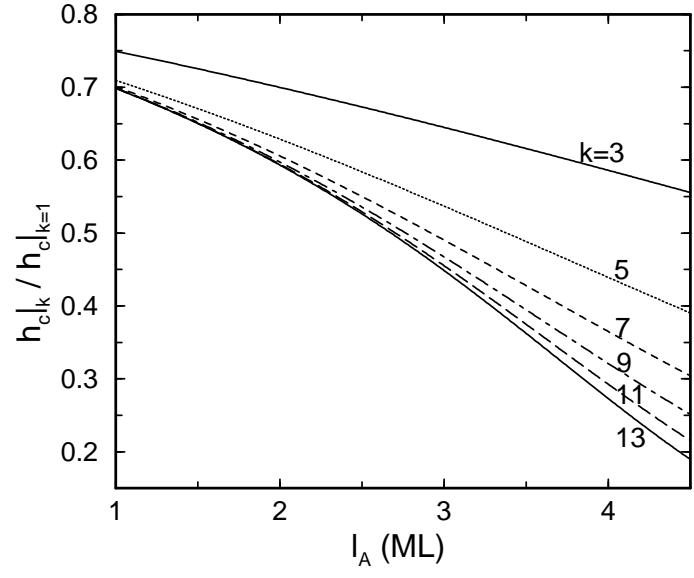}}
\caption{\label{fig-hc-l}Rescaled kinetic critical thickness $h_c|_k
/h_c|_{k=1}$ as a function of average strained layer thickness $l_A$,
for successive strained layer numbers. $l_B=8.5$ nm, $v_A=1$ ML/min,
and $T=550$ $^{\circ}$C.}
\end{figure}

To further study the effect of layer thickness, we plot the rescaled
critical thickness $h_c|_k/h_c|_{k=1}$ as a function of spacer layer
thickness $l_B$ in the inset of Fig. \ref{fig-hc-k}, for different 
layer numbers $k=3$, $5$, $7$, and $11$. $h_c|_k$ increases as the 
spacer thickness $l_B$ increases and the layers number $k$ decreases, 
in agreement with Fig. \ref{fig-hc-k} and recent Ge/Si(001) 
experiments. \cite{lethanh00} When $l_B$ is large enough, 
the increase of $h_c|_k$ becomes much slower, with the value 
approaching that of the single layer growth, as analyzed in Sec. 
\ref{sec:spacer-A}. The relationship between the rescaled $h_c|_k$
and the thickness $l_A$ of buried strained layers is presented in 
Fig. \ref{fig-hc-l}, which shows the opposite effect of $l_A$. 
Larger $l_A$ causes rougher morphology of buried interfaces, and 
hence the enhanced perturbation on surface as a result of the strain 
fields coupling [see also Eq. (\ref{eq-h-hat})]. Consequently, for 
larger $l_A$ instability occurs earlier, with smaller value of 
$h_c|_k$.

\begin{figure}
\resizebox{9.cm}{!}{\includegraphics{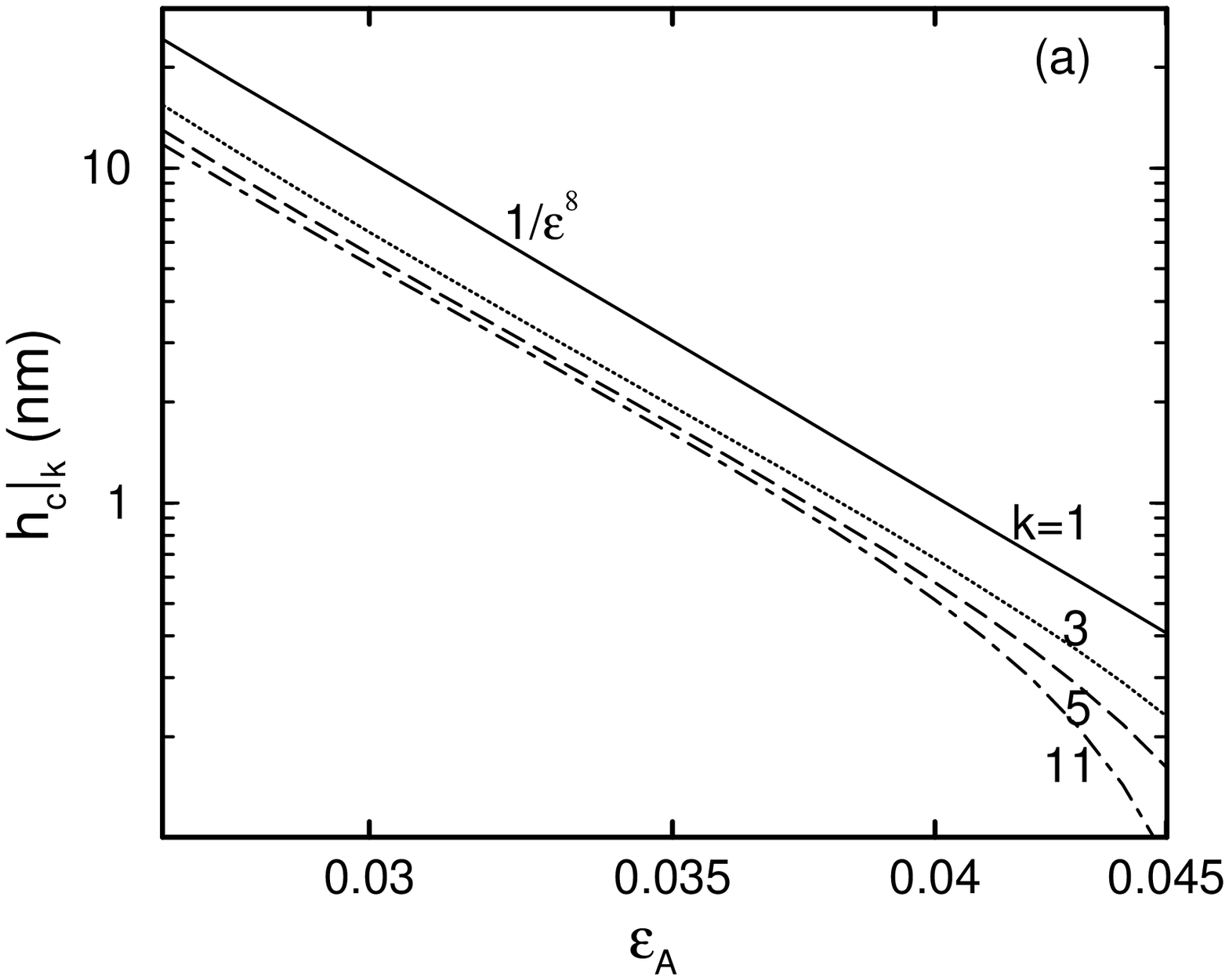}}
\vskip 0.5cm
\resizebox{9.cm}{!}{\includegraphics{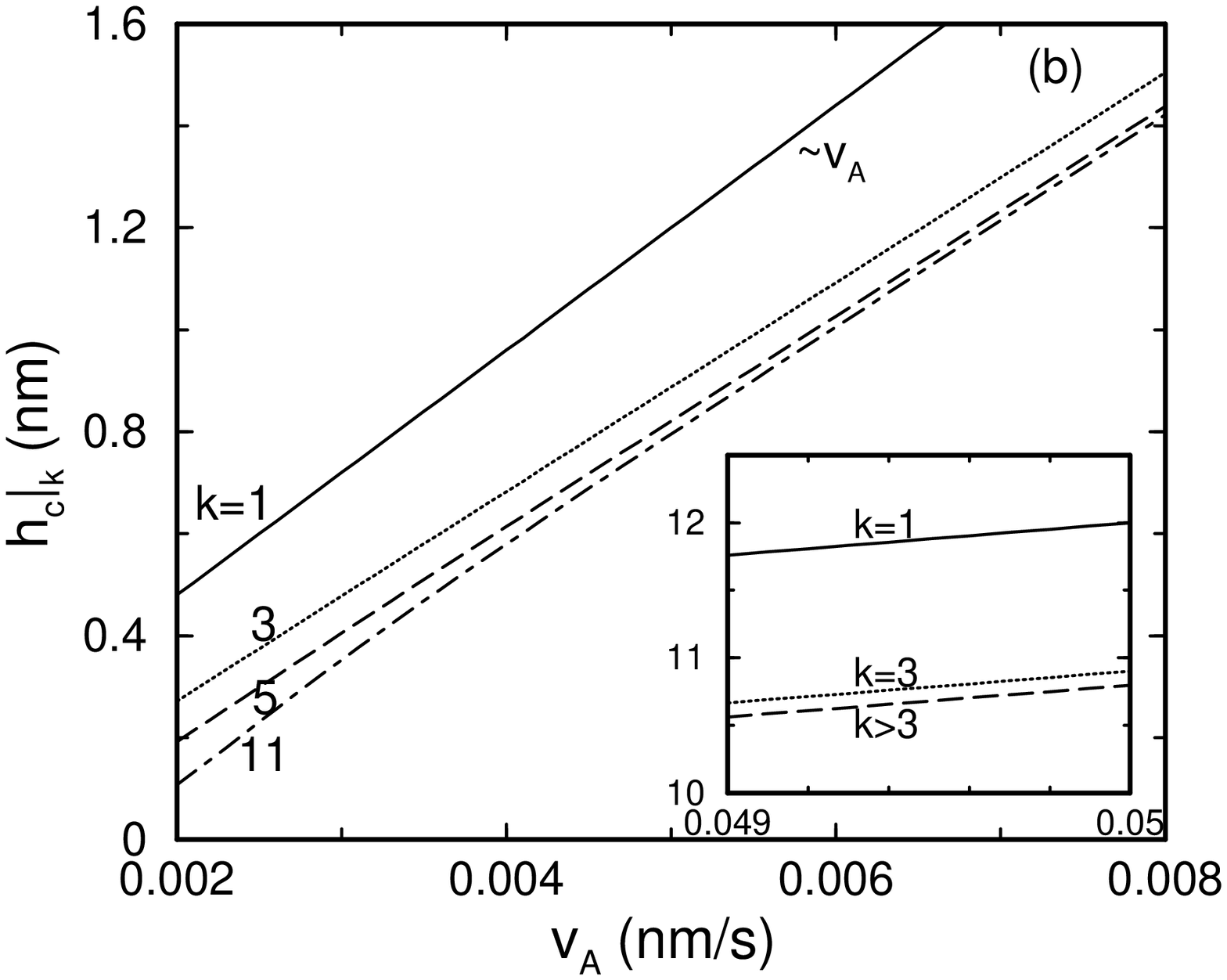}}
\caption{\label{fig-hc-epsv}Kinetic critical thickness $h_c|_k$ as a 
function of (a) misfit $\epsilon_A$ and (b) deposition rate $v_A$, for 
different layer numbers $k=1$, $3$, $5$, and $11$. The parameters are
$l_A=4$ ML, $l_B=9$ nm, and $T=550$ $^{\circ}$C for both (a) and (b),
as well as $v_A=1$ ML/min for (a). Note the inset in (b) where the 
range of $v_A$ is almost ten times larger.}
\end{figure}

The behavior of kinetic critical thickness $h_c|_k$ with respect 
to misfit strain $\epsilon_A$ and deposition rate $v_A$ is shown
in Fig. \ref{fig-hc-epsv} (a) and (b). The results of $k=1$ there
correspond to those of single-layer growing film with single 
component, for which the kinetic critical thickness is found to 
follow the $\epsilon_A^{-8}$ power law and increase linearly with 
$v_A$. \cite{spencer91} Similar properties can also be obtained 
in our calculations for strained/spacer multilayers with layer 
number $k>1$ and at small misfit [see the range $\epsilon_A < 
4\%$ in Fig. \ref{fig-hc-epsv} (a)] and large deposition rate 
[see Fig. \ref{fig-hc-epsv} (b) and its inset]. However, small 
deviations do occur for large $\epsilon_A$ and very small $v_A$, 
as illustrated in Fig. \ref{fig-hc-epsv}, especially for large 
$k$ systems. Around and beyond the misfit of $4\%$, Fig. 
\ref{fig-hc-epsv} (a) shows that $h_c|_k$ for $k>1$ decreases
faster than $\epsilon_A^{-8}$ as misfit increases, with the 
behavior deviated from the power law; also, as presented in
Fig. \ref{fig-hc-epsv} (b), the linear relationship between 
$h_c|_k$ and $v_A$ is deviated slightly at very small deposition 
rate. These deviations are more obvious for larger number $k$. 
But when $k$ is large enough ($k\geq 11$ for the parameters
of Fig. \ref{fig-hc-epsv}), the graphs of $h_c|_k$ for different 
$k$ are almost identical in Fig. \ref{fig-hc-epsv} (a) and 
(b), due to the convergence effect obtained analytically in 
Eq. (\ref{eq-Ak-infty}) and shown numerically in Fig. \ref{fig-hc-k} 
for large $k$.

Note that the \textit{kinetic} critical thickness $h_c|_k$ studied
here, which is for the onset of morphological instability 
\textit{during layer growth}, is different from the wetting layer 
thickness $h_{\rm WL}$ which corresponds to \textit{equilibrium} 
critical thickness \cite{sunamura95} (\textit{postgrown} and 
thermodynamically stable) for the 3D island formation. Usually 
$h_c|_k > h_{\rm WL}$ due to the consumption of wetting layer in 
the formation of 3D islands. E.g., in the molecular-beam epitaxy
(MBE) experiment of single layer Ge on Si(100), \cite{sunamura95} 
$h_c|_{k=1} = 3.7$ ML at temperature $T=700$ $^{\circ}$C, while 
$h_{\rm WL}=3$ ML.

For this stacked strained/spacer multilayer, due to the presence 
of spacer layers which are of the substrate material, the 
mechanism for morphological instability is more analogous to that 
of single-layer film growth compared to the tensile/compressive
structure. But the crucial factor is still the coupling of strain
fields in different strained layers. Also,
from the general expressions in Sec. \ref{sec:general}, it is 
straightforward to obtain the results of morphological perturbation 
for the case of nonplanar and rough spacer layer $B$, that is, 
$\epsilon_i=0$ but $\hat{h}_i \neq 0$ for even number $i$. The
corresponding stability properties and behavior of kinetic critical
thickness are more complicated than those of the system with planar 
spacer layers studied above, due to the rough strained/spacer 
interfaces leading to a stronger coupling of strain fields.

\section{\label{sec:concl}Conclusions}

We have studied the early morphological evolution and stability/%
instability properties of coherent multilayer structures with the
stacking of layers of different materials. Generally, the diffusion 
in the bulk film is assumed to be negligible, thus the phenomena 
observed in the whole multilayer films, such as the lateral 
composition modulations in tensile/compressive multilayer and the
vertically correlated arrays of islands in stained/spacer system,
are in fact surface-induced and arise through the stacking of layers 
with the coupling of strain fields between surface and underlying 
layers. We have proposed a recursive procedure to directly 
calculate the elastic fields of a multilayer system to first order
of perturbations. Then, based on the surface diffusion processes,
the evolution equation of surface morphological perturbation is
expressed through the elastic and morphological properties of all 
the layers underneath, including the misfit strains and buried 
interface perturbations. From the exact solution of this 
first-order equation, which is in terms of material parameters
and growth conditions for all the deposited layers, we have
determined the properties of stress-driven morphological 
instability for both the tensile/compressive and strained/spacer 
multilayers.

In this paper, we have only applied our results of early morphological 
evolution, derived in sections \ref{sec:model} and \ref{sec:general}, 
to two kinds of periodic multilayer structures, but more complicated
structures can be studied similarly. Note that for multilayers
with very thin layer thickness, i.e., for short-period superlattices,
although our results can qualitatively explain and predict some 
experimental observations, more realistic factors such as the 
wetting effect should be included in our model to quantitatively 
describe the system properties. Also, studies beyond the linear 
analysis are needed to determine the details of morphological 
modulations or islands evolution during the multilayer growth.
All these theoretical results can be useful in the fabrication 
of high-quality and novel multilayer structures.

\begin{acknowledgments}
This work was supported by the NSERC of Canada.
\end{acknowledgments}

\appendix*

\section{\label{append} Results for first and second layer perturbations}

The morphological perturbation $\hat{h}_1$ for first deposited layer 
$1$ is the same as that derived in the case of single-layer film 
growth. For the planar film-substrate interface ($\zeta=0$), it is
\begin{equation}
\hat{h}_1(q,t)=e^{\sigma_1 t} \hat{h}^0(q),
\label{eq-h1-t}
\end{equation}
with the assumption of identical elastic constants. \cite{asaro72} 
Here $\sigma_1=\Gamma_1 (E'\epsilon_1^2 q^3 - \gamma_1 q^4)$ and
$\hat{h}^0(q)$ is the initial perturbation. For the layer with
fixed thickness $l_1$, Eq. (\ref{eq-h1-t}) leads to
\begin{equation}
\hat{h}_1(q)=e^{\sigma_1 l_1/v_1} \hat{h}^0(q).
\label{eq-h1-l}
\end{equation}
The perturbation $\hat{h}_2$ for second deposited layer can be
calculated from the method in Sec. \ref{sec:model} and the expression
(\ref{eq-h1-l}) for $\hat{h}_1$, with the results
\begin{eqnarray}
\hat{h}_2(q,t) &=& \frac{1}{qv_2 + \sigma_2} \left [ \left (
qv_2+ \Gamma_2 \left ( E' \epsilon_2\epsilon_1 q^3 - \gamma_2 q^4
\right ) \right ) e^{\sigma_2 t} \right. \nonumber\\
&+& \left. \Gamma_2 E' \epsilon_2 (\epsilon_2 - \epsilon_1) 
q^3 e^{-q v_2 t} \right ] e^{\sigma_1 l_1/v_1} \hat{h}^0(q),
\label{eq-h2-t}
\end{eqnarray}
where $\sigma_2=\Gamma_2 (E'\epsilon_2^2 q^3 - \gamma_2 q^4)$, and
\begin{equation}
\hat{h}_2(q) = \hat{h}_2(q,t=l_2/v_2)
\label{eq-h2-l}
\end{equation}
when second layer has fixed thickness $l_2$.

In some real growth systems, \cite{lafontaine98-2,tiedje01} the 
substrate surface is nonplanar, corresponding to the case of
rough ($\zeta \neq 0$) film-substrate interface. The elastic
state of both film and substrate is altered by this rough interface,
with the detailed results shown in Ref. \onlinecite{huang02}. Here
we use these elastic results to obtain the expression of first
layer perturbation for single-component film:
\begin{equation}
\hat{h}_1(q,t) = \frac{1}{qv_1 + \sigma_1} \left [
\left ( qv_1 - \Gamma_1\gamma_1 q^4 \right ) e^{\sigma_1 t}
+ \Gamma_1 E' \epsilon_1^2 q^3 e^{-qv_1 t} \right ] \hat{\zeta}(q),
\label{eq-h1-tz}
\end{equation}
where we have assumed the initial condition $\hat{h}_1(q,t=0)=
\hat{\zeta}(q)$. Note that the contribution of nonzero interface
morphology $\zeta$ is reflected in the second term of r.h.s. of
Eq. (\ref{eq-h1-tz}), which is exponentially damped as the film 
grows. Therefore, the occurrence of instability is still determined
by the real part of perturbation growth rate $\sigma_1$, as in
the case of flat film-substrate interface [see Eq. (\ref{eq-h1-t})],
\cite{asaro72,spencer91,guyer95,spencer00,leonard98,huang02} 
and the introducing of nonzero $\zeta$ does not influence the
stability/instability condition for single-layer film growth. 
On the other hand, Eq. (\ref{eq-h1-tz}) shows that compared to
the case of flat interface $\zeta=0$, the kinetic critical thickness 
for the onset of 3D instability \cite{spencer91} (see also Sec. 
\ref{sec:spacer} for the definition) decreases for rough interface
($\zeta\neq 0$). Also, the nonzero substrate roughness $\zeta$ can 
affect the detailed film morphology, as found in experiments. 
\cite{lafontaine98-2,tiedje01}

The second layer surface perturbation can then be calculated for
rough film-substrate interface, with the result:
\begin{eqnarray}
\hat{h}_2(q,t) &=& \frac{1}{qv_2 + \sigma_2} \bigglb \{
\frac{1}{qv_1 + \sigma_1} \left [ \left ( qv_1 - \Gamma_1\gamma_1 q^4 
\right ) e^{\sigma_1 l_1/v_1} + \Gamma_1 E' \epsilon_1^2 q^3 e^{-q l_1} 
\right ] \nonumber\\
&\times& \left [ \left ( qv_2+ \Gamma_2 \left ( E' \epsilon_2\epsilon_1 
q^3 - \gamma_2 q^4 \right ) \right ) e^{\sigma_2 t} + \Gamma_2 E' 
\epsilon_2 (\epsilon_2 - \epsilon_1) q^3 e^{-q v_2 t} \right ] 
\nonumber\\
&+& \Gamma_2 E' q^3 \epsilon_2 \epsilon_1 \left ( e^{-qv_2 t} - 
e^{\sigma_2 t} \right ) e^{-q l_1} \biggrb \} \hat{\zeta}(q).
\label{eq-h2-tz}
\end{eqnarray}
Similar to Eqs. (\ref{eq-h1-l}) and (\ref{eq-h2-l}), the surface
morphological perturbations of first and second layer with fixed
layer thicknesses, i.e., $\hat{h}_1(q)$ and $\hat{h}_2(q)$,
can be obtained by substituting $t=l_1/v_1$ and $t=l_2/v_2$ into
Eqs. (\ref{eq-h1-tz}) and (\ref{eq-h2-tz}), respectively.


\newpage

\end{document}